# Steering interfacial molecular structures by substrate-molecule charge transfer

Benjamin Achatz,*[a] Matthias Zeilerbauer,[a] Marco Thaler,[a] Milan Ončák,[b] Shadi Fatayer,[c] and Laerte L. Patera*[a]

a. Department of Physical Chemistry, University of Innsbruck, 6020 Innsbruck (Austria)
E-mail: benjamin.achatz@uibk.ac.at, laerte.patera@uibk.ac.at

b. Department of Ion Physics and Applied Physics, University of Innsbruck, 6020 Innsbruck (Austria)

c. Physical Science and Engineering Division (PSE), King Abdullah University of Science and Technology (KAUST), Thuwal 23955-6900 (Saudi Arabia)

Intermixed self-assembled monolayers (SAMs) provide a powerful approach for tuning inorganic-organic interfaces, but their local structure and energetics often remain elusive. Here, an intermixed stripe phase of fullerene $C_{60}$ and Zn(II)-5,10,15,20-tetrakis(4-aminophenyl)porphyrin (Zn-TAPP) on a Ag(111) surface is identified, establishing a direct link between molecular assembly and charge transfer. Scanning tunnelling spectroscopy (STS) and Kelvin probe force spectroscopy (KPFS) have been used to map electronic states and local contact potential differences (LCPD). Compared to close-packed $C_{60}$ islands, the stripe phase exhibits a larger interfacial dipole and increased occupation of previously unoccupied $C_{60}$ states, consistent with enhanced charge transfer from Ag(111) to $C_{60}$. Analysis indicates that larger $C_{60}$ spacing in the stripe geometry reduces Coulomb repulsion, stabilizing the intermixed phase and enhancing charge transfer with respect to close-packed domains. These results provide direct evidence of charged $C_{60}$ molecules in stripe structures and guidelines for engineering intermixed SAMs with tailored interfacial electronic properties.

## Introduction

Organic electronic devices have emerged as a promising alternative to conventional inorganic semiconductor technologies. Their appeal lies in the potential for cost and energy efficient fabrication, including solution processing and low temperature deposition.[1,2] The chemical tunability of organic molecules enables devices with tailored electronic and optical properties across a wide range of applications.[1,3] Device performances have been improved by tuning both the molecular components[4] and the metal electrodes,[5] with particular attention on interfacial energetics and charge injection barriers at the metal–organic interface.[6] Among various strategies to control this interface, the usage of self-assembled monolayers (SAMs) has proven especially powerful.[7–9] Intermixed SAMs allow for precise modulation of work function, local dipoles, and contact resistance, thereby improving device efficiency[10,11] and transport properties[12]. Realizing this potential requires a molecular level understanding of the intermolecular interactions and packing rules that govern the formation of such intermixed SAMs.

Investigations of SAMs and metal–organic interfaces have relied extensively on surface-sensitive spectroscopic techniques such as X-ray photoelectron spectroscopy (XPS) and ultraviolet photoelectron spectroscopy (UPS).[11,13,14] These methods enable the determination of composition, bonding, level alignment, and average molecular orientation. However, these techniques inherently perform spatial averaging over the illuminated area, thereby masking the molecule-scale heterogeneity and local physicochemical properties that are intrinsic to intermixed SAMs.[15] Conversely, scanning probe microscopy (SPM) techniques enable direct, local interrogation of structure and energetics at the nanometre scale.[16,17] Scanning tunnelling microscopy and spectroscopy (STM/STS) provide molecular scale information on packing, defects, and electronic states of SAMs.[18–20] Non-contact atomic force microscopy (nc-AFM) can be exploited to precisely investigate the chemical structure[21], adsorption configurations[21,22] and adsorption height with sub-angstrom resolution.[23,24] Additionally, Kelvin probe force spectroscopy (KPFS) probes local contact potential differences (LCPD), enabling the mapping of local surface dipoles with sub-nanometre scale resolution.[25,26]

The realization of intermixed phases is of special interest for optoelectronic devices based on electron donor-acceptor (D-A) pairs,[27] where, upon photoexcitation, charge separation occurs at the heterojunctions.[28] Device efficiency can be enhanced by increasing the interfacial area between donor and acceptor domains[29] and controlling the local structure of the interfaces.[28] Metalloporphyrins and $C_{60}$ are commonly used as an electron D-A pair due to the formation of stabilized charge-separated states.[30,31] Intermolecular interaction in bulk heterostructures of porphyrins and $C_{60}$ are dominated by strong van der Waals interactions.[31,32]

Intermixed SAMs of prototypical D-A pairs such as Zn-porphyrins and $C_{60}$ have been reported on a Ag(110) surface, where coverage dependent pore and stripe phases emerge.[33,34] Stripe formation of $C_{60}$ on a graphite surface has previously been attributed to the interplay between attractive short-range Van der Waals interactions and long-range dipole-dipole repulsion between charged $C_{60}$ molecules,[35] similar to the behaviour reported for rubrene chains on Au(111).[36] However, direct experimental evidence for molecular charging within these $C_{60}$ stripe phases has so far been lacking.

Here, we report the on-surface synthesis of an intermixed stripe phase of $C_{60}$ and Zn(II)-5,10,15,20-tetrakis(4-aminophenyl)porphyrin (Zn-TAPP) on the close packed Ag(111) surface. We show that the combination of KPFS and STS enables the experimental investigation and comparison of the degree of substrate-molecule charge transfer for $C_{60}$ in close-packed islands and in the intermixed stripe phase. KPFS provides insights into variations of the local interfacial dipoles, resulting from substrate-molecule charge transfer.[37,38] STS can be used to elucidate relative shifts in unoccupied-states near the Fermi level ($E_F$), consistent with partial charge transfer to $C_{60}$.[39,40] Linking the experimentally determined degree of charge transfer to the structure of the distinct phases provides evidence for the crucial role of electrostatics in the formation of these intermixed SAMs on metal surfaces.

## Results

### Assembly transformation upon annealing

A sub-monolayer of $C_{60}$ and Zn-TAPP molecules has been obtained by deposition on a clean Ag(111) held at 300 K. Post-annealing up to 415 K for 2 h results in the phase separation into large domains of $C_{60}$ and Zn-TAPP (see Fig. 1a). The observed phase separation is consistent with the strong Van der Waals interactions reported for $C_{60}$ molecular crystals, making segregation favourable.[41] On closed-packed transition metal surfaces $C_{60}$ forms hexagonal islands,[42–45] which nucleate primarily at step edges.[46] Fig. 1b displays the characteristic bright and dim contrasts of $C_{60}$ molecules adsorbed on Ag(111).[42–45] The dim contrast has been attributed to the formation of single-atom vacancies underneath the respective $C_{60}$ molecules, steered by annealing to 415 K.[42–46] STM images indicate that approximately 33% of $C_{60}$ molecules adsorb on reconstructed sites (i.e. dim configuration). On Ag(111), $C_{60}$ form a (2√3x√3)R30° superstructure (Fig. S1) with centre-to-centre distances of 1.00 nm.[44] The STM appearance of individual $C_{60}$ molecules reflects their specific adsorption geometry:[43–45,47,48] adsorption on a 6-membered ring results in 3-fold symmetry and adsorption on a bond between two 6 membered rings leads to 2-fold symmetry.[47] Ordered Zn-TAPP domains are observed to form preferentially at the centre of Ag(111) terraces. Fig. 1c presents a two-dimensional rectangular lattice of Zn-TAPP (lattice parameters: a = 1.68 nm, b = 1.61 nm, α = 87°).

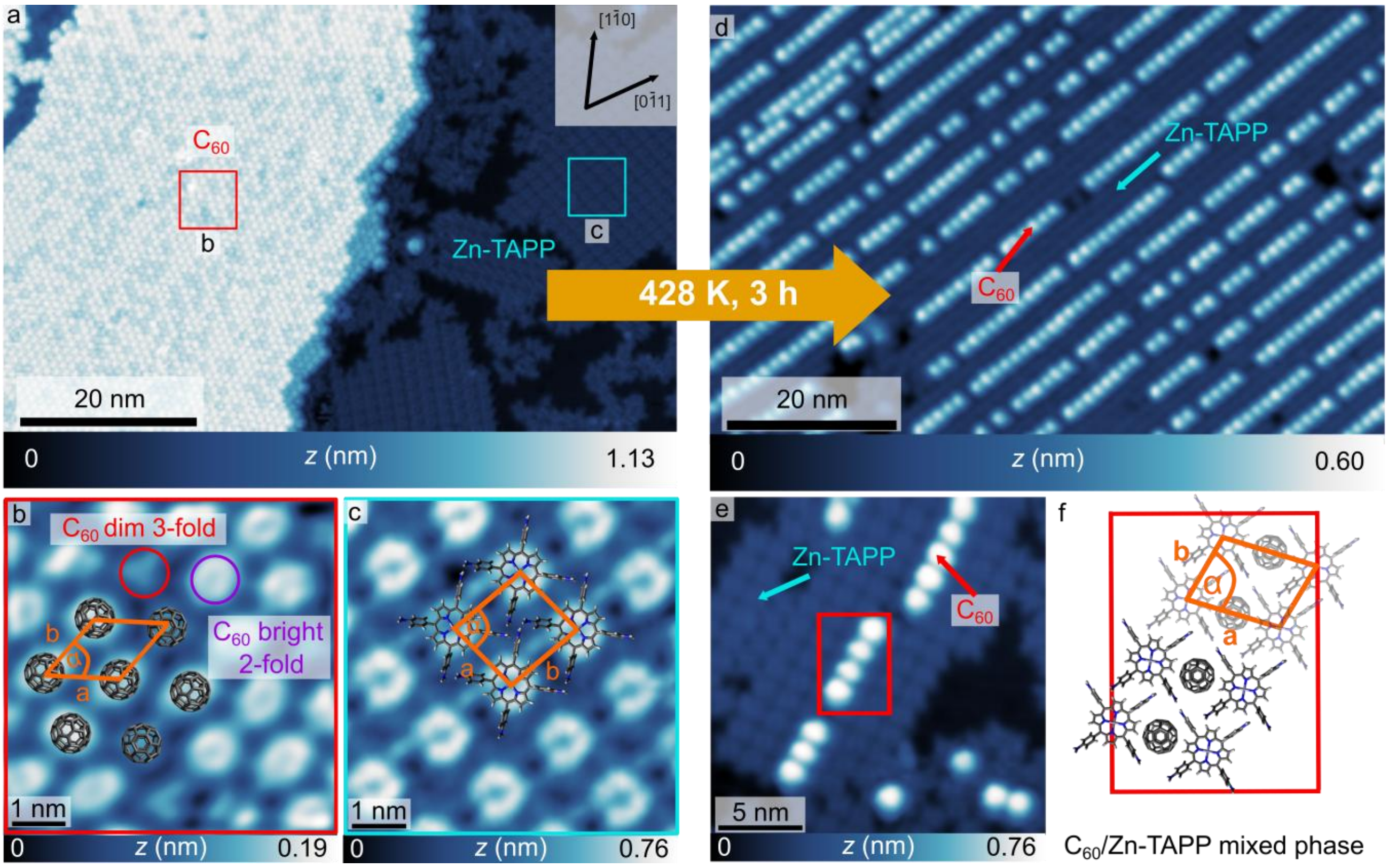

**Fig. 1.** Formation of an intermixed self-assembled monolayer on Ag(111). (a) Constant-current STM image acquired upon deposition of $C_{60}$ and Zn-TAPP on Ag(111) and subsequent annealing to 415 K for 2 h. (b-c) High-resolution STM images of ordered Zn-TAPP and $C_{60}$ phases, respectively, overlayed with structural models. (d-e) Constant-current STM images acquired after annealing to 428 K for 3 h. (f) Model of the intermixed phase of $C_{60}$ and Zn-TAPP. Unit cells are indicated in orange. Measurement parameters: (a) tunneling current $I$ = 20 pA, sample bias voltage $V$ = 0.5 V. (b-c) $I$ = 10 pA, $V$ = 0.5 V. (d) $I$ = 5 pA, $V$ = 1.0 V. (e) $I$ = 10 pA, $V$ = 0.6 V.

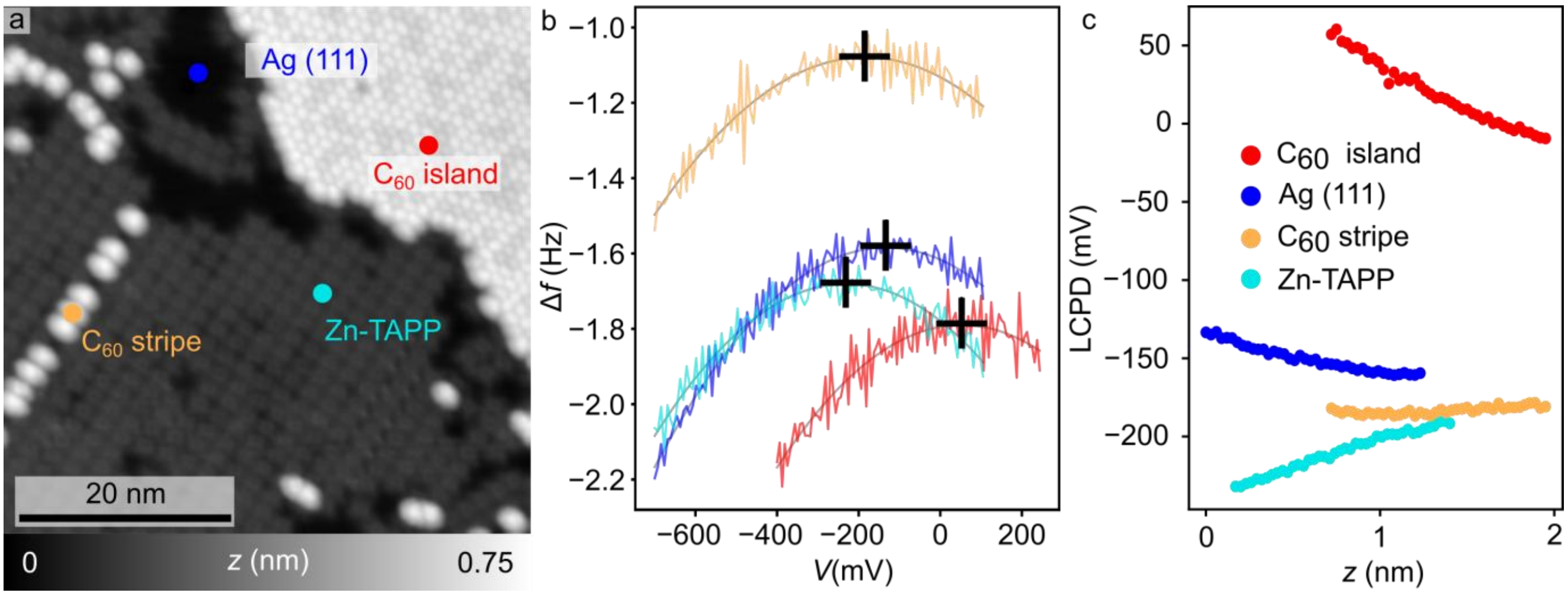

**Fig. 2.** Height-dependent KPFS. (a) Constant-current STM image acquired after annealing to 428 K for 3 h. (b) KPF spectra acquired at the positions marked in (a): on a close packed $C_{60}$ island (red), on a $C_{60}$ in a one-dimensional stripe (orange), on a Zn-TAPP island (cyan) and on Ag(111) (blue). Parabolic fits and corresponding parabola peaks are indicated for each spectrum. (c) Fitted LCPD plotted as a function of the tip height $z$. The LCPD($z$) curves were offset by the corresponding apparent heights obtained from (a). Measurement parameters: (a) $I$ = 20 pA, $V$ = 0.5 V. (b) Feedback opened on the marked spots at $I$ = 5 pA, $V$ = 0.6 V, $z$ = 0.3 nm. (c) Feedback opened on the marked positions at $I$ = 5 pA, $V$ = 0.6 V. Positive $z$ values correspond to a tip height increase compared to the STM setpoint. AFM oscillation amplitude $A$ = 0.3 nm.

Annealing to 428 K for 3 h results in the formation of an intermixed phase (Fig. 1d, e), where $C_{60}$ molecules arrange in one-dimensional stripes constrained by a surrounding ordered Zn-TAPP array. The structural model of the intermixed phase shown in Fig. 1f reveals pores formed between four Zn-TAPP molecules, with each pore accommodating a single $C_{60}$ (unit cell parameters: a= 2.04 nm, b= 1.47 nm, α = 77°), similar to the structure reported on Ag(110).[46] The extracted intermolecular distances of $C_{60}$ for the intermixed phase (1.47±0.02 nm) are considerably larger than those in the close-packed phases (1.00±0.03 nm). Several stripe orientations are observed (see Fig. S2). These orientations arise from the presence of differently oriented Zn-TAPP domains on the Ag(111) substrate. The orientational distribution of the intermixed phase is therefore mainly associated with the domain structure and molecular packing of the Zn-TAPP template.

### Kelvin Probe Force Spectroscopy

For isolated $C_{60}$ molecules adsorbed on an unreconstructed Ag(111) surface, density functional theory (DFT) calculations predict charge transfer of up to 0.5 electrons per molecule from Ag to $C_{60}$.[49] On reconstructed sites, due to the larger hybridization with the substrate, an even larger degree of charge transfer can be expected.[39,44] The resulting negative charging of the $C_{60}$ molecules introduces significant intermolecular repulsion, potentially disfavouring dense packing and instead promoting the formation of energetically favourable linear arrangements. To elucidate this point, KPFS has been performed to probe the degree of substrate-$C_{60}$ charge transfer. Individual spectra have been acquired on top of $C_{60}$ molecules both in the islands and in the intermixed phase, as well as on Zn-TAPP molecules and on the Ag(111) substrate (see Fig. 2a). A parabolic behaviour of the frequency shift ($\Delta f$) upon sweeping the sample bias voltage ($V$) is observed for all spectra in Fig. 2b. The sample bias voltage corresponding to the maximum of the parabola ($V^*$) is defined as the local contact potential difference (LCPD). For $V$ = LCPD, the electrostatic contribution to the force gradient is minimized. As a result, the corresponding $\Delta f$ value ($\Delta f^*$) mostly reflects Van der Waals contributions.[25,26,50]

In KPFS, the LCPD value is influenced both by the atomic structure of the tip apex as well as by the tip–sample separation, owing to spatial averaging effects.[25,26] Consequently, meaningful comparisons of LCPD shifts must be carried out using the same probe and constant tip–sample distances. To disentangle electrostatic effects from height-related artefacts, KPFS measurements should be performed at several tip heights.[25,26]

To examine the LCPD dependence on the tip height ($z$), the tip was stabilized at a defined STM setpoint, the feedback was disabled, and the tip was retracted in discrete height steps. The LCPD($z$) curve measured above the $C_{60}$ island exhibits a more pronounced increase in LCPD compared to the one acquired on Ag(111). This effect indicates a larger local surface dipole on the $C_{60}$ islands than on the substrate, being induced by an accumulation of negative charge on the $C_{60}$ islands, consistent with previous DFT studies.[49] In contrast, the lower LCPD($z$) values observed for the Zn-TAPP domain compared to Ag(111) is attributed to the push-back effect, which suppresses electron spill-out at the interface.[51–53] Notably, the LCPD($z$) curve for the linear $C_{60}$ stripe is rather flat, suggesting a smaller surface dipole compared to the $C_{60}$ island. These trends are also observed when plotting the LCPD against $\Delta f^*$ (see Fig. S4). Fig. 3a shows a $C_{60}$ island adjacent to a $C_{60}$ stripe. The corresponding KPFS map (Fig. 3b) indicates that the LCPD of the $C_{60}$ molecules in the stripe phase is lower than the value found for the $C_{60}$ islands.

It is important to note that an unambiguous assignment of molecular charge states based on LCPD measurements on metal-supported systems is nontrivial. The reduction of the surface dipole due to the push-back effect can partially mask the LCPD increase expected from negative charge accumulation.[54] Nevertheless, a qualitative comparison remains possible, particularly by considering trends in the LCPD($z$) curves. In this context, the KPFS data shown in Fig. 2 and 3 suggest a pronounced negative charge accumulation for molecules within the islands, whereas those in the stripe phase seem to exhibit only marginal, if any, charge transfer. However, since the $C_{60}$ islands and stripe phases differ in their lateral dimensions, the lower measured LCPD values for the stripes may also be influenced by the finite lateral resolution of KPFM, which leads to spatial averaging effects.[55]

To better address the effect of dimensionality on the KPFS contrast, we acquired spectra along the Zn-TAPP–$C_{60}$ island interface and across a $C_{60}$ stripe (Fig. 4a), enabling an evaluation of the spatial resolution of the KPFS measurements. As shown in Fig. 4b, the island region exhibits a larger absolute LCPD value. However, the boundary between the $C_{60}$ and Zn-TAPP phases appears significantly smeared out. When considering the LCPD profile across a $C_{60}$ molecule in a stripe, the full width at half maximum (FWHM) is broader (2.1 nm) than the actual size of a $C_{60}$ molecule (≈ 0.7 nm). This indicates that the KPFS spatial resolution is insufficient to resolve the intrinsic LCPD of the one-dimensional stripes accurately, as substantial averaging with the surrounding Zn-TAPP layer occurs. The lateral resolution of KPFS also further decreases with increasing tip-height as shown for the edge of a $C_{60}$ island in Fig. S5. Overall, the averaging effects reduce the contrast on the one-dimensional intermixed phase, underrepresenting their actual local charges. To attempt to retrieve the actual LCPD value of $C_{60}$ in the stripe phase, we adopt the concept of a point spread function (PSF).[56] By attributing a gaussian PSF to the KPFS measurements shown in Fig. 4b, the measured LCPD contrast can be deconvolved (see Fig. S6 and S7). This analysis suggests that the LCPD contrast of the $C_{60}$ in the stripe phase is underestimated by a factor of 7.8±0.6, suggesting that the actual degree of charge transfer for the $C_{60}$ molecules in the stripe phase is significantly larger than directly inferred from the measured LCPD contrast (see Fig. S9).

Notably, the LCPD profiles in Fig. 4b show no evidence of positive charge located at the Zn-TAPP molecules at either interface with $C_{60}$, indicating that the negative charge accumulation observed on $C_{60}$ is likely due to charge transfer from the substrate rather than from neighbouring Zn-TAPP molecules.

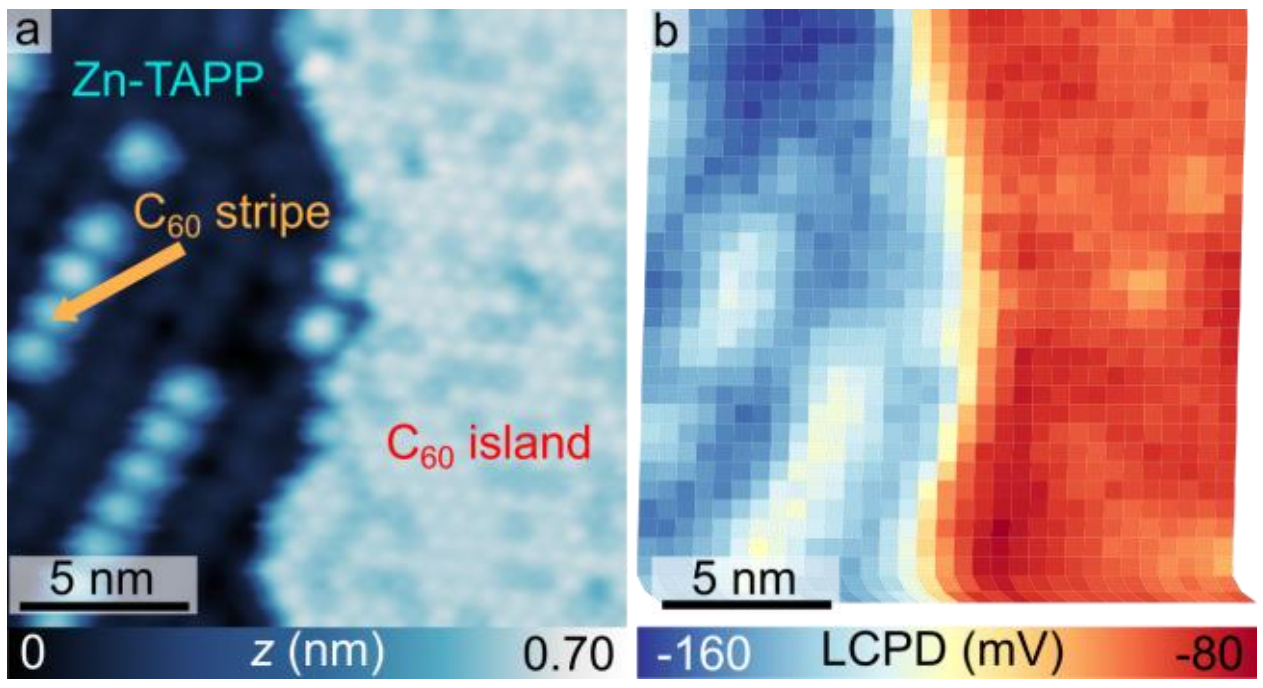

**Fig. 3.** KPFS mapping. (a) Constant-current STM image and (b) corresponding drift-corrected LCPD map of a $C_{60}$ island (right) and the intermixed $C_{60}$/Zn-TAPP phase (left). Measurement parameters: (a, b) $I$ = 5 pA, $V$ = 0.6 V. (b) $z$ = 0.3 nm, $A$ = 0.3 nm.

### Δ$f$($z$) spectroscopy measurements

Site-specific $\Delta f(z)$ spectroscopy (Fig. 4c) has been performed to determine relative height differences between the $C_{60}$ molecules in the distinct phases on Ag(111). The relative shifts of the tip height $z$ corresponding to the turnaround point of the $\Delta f(z)$ signal ($z^*$) provide the height differences. Relative height difference ($\Delta z$) of around 0.08 nm between the dim 3-fold and bright 2-fold molecules can be attributed to adsorption of the dim molecules on reconstructed sites.[43,45] While $\Delta z$ values were found to be slightly tip dependent, varying between 0.05 and 0.10 nm, the qualitative trends were consistent for different tips. The spectra shown in Fig. 4c were likely obtained with a $C_{60}$ functionalized tip.[53] Notably, $\Delta f(z)$ curves recorded on $C_{60}$ molecules in the intermixed phase exhibit a $z^*$ value close to the one of the dim 3-fold sites, suggesting adsorption on reconstructed sites.

### Characterization of electronic properties with tunnelling spectroscopy

Tunnelling conductance (d$I$/d$V$) spectra acquired for $C_{60}$ molecules in the islands (Fig. 5a) and in the intermixed phase (Fig. 5b) exhibit electronic states close to the Fermi level ($E_F$), attributed to the lowest unoccupied molecular orbitals (LUMOs) of the $C_{60}$ molecules (Fig. 5c-e).[47] While gas-phase $C_{60}$ exhibits three-fold degenerate LUMOs,[46,57] adsorption on metal substrates results in the degeneracy lifting due to hybridization.[44,57] Distinct peak shapes of the $C_{60}$ LUMO in Fig. 5c-f can be thus attributed to peak splitting and broadening stemming from different degrees of hybridization with the underlying substrate, as well as from π-π orbital hybridization between neighbouring $C_{60}$ molecules,[44] consistent with observations in thin-film systems.[58]

The apparent degree of charge transfer has been estimated by determining the area of the LUMOs lying below $E_F$ (see below). The d$I$/d$V$ spectra were fitted with three Gaussian functions for the LUMOs[57] and an additional Gaussian for the onset of the LUMO+1 (see SI, Table S1). To determine the total areas of the LUMO orbitals the contributions of the three fitting components (between $V_{min}$ = -0.2 V and $V_{max}$ =1.2 V) were summed up. The degree of charge transfer ($n_{e^-}$) can be estimated by calculating the apparent LUMO occupation, assuming full occupation of the three LUMOs corresponds to six electrons per molecule (see SI Table S2):

$$n_{e^-} = 6 \times \frac{\int_{V_{min}}^{0} dI/dV}{\int_{V_{min}}^{V_{max}} dI/dV} \quad (1)$$

Minimal crossing of $E_F$ in spectra of bright 3-fold and 2-fold $C_{60}$ indicates negligible charge transfer (0.04 and 0.03 electrons per molecule, respectively). A slightly higher value is calculated for dim 3-fold $C_{60}$ molecules (0.08 electrons per molecule), although this still corresponds to a relatively low level of charge accumulation. In contrast, an increased degree of charge transfer of 0.50 electrons per molecule was determined for $C_{60}$ molecules in the intermixed phase, suggesting that the one-dimensional phase enables a larger degree of charge transfer with respect to the dense phase.

Interpretation of energy shifts in d$I$/d$V$ spectra requires a detailed understanding of the energy alignment at the metal-organic interface.[57,59,60] It is important to note that, for molecules adsorbed on metal surfaces, d$I$/d$V$ spectroscopy do not provide the actual energy alignment of frontier orbitals. In electron spectroscopies, charge injection and removal are associated with an additional energy cost, commonly referred to as the Hubbard on-site energy, which leads to an apparent widening of the HOMO–LUMO gap for isolated molecules (see $C_{60}$ (gas) in Fig. 6).[57] Upon adsorption on a metallic substrate the injected charge is screened by an image charge within the substrate ($\Delta E_{pol.(sub.)}$ in Fig. 6), reducing the Hubbard on-site energy.[57] Moreover, a reduction of the surface dipole caused by the push-back effect results in a downwards shift of the vacuum level and of all molecular orbitals ($\Delta E_{p.b.}$ in Fig. 6).[52] The energy alignment of the frontier orbitals is further governed by hybridization with the substrate, which depends on the substrate reactivity, the adsorbate electronic structure, and its adsorption height.[52] In d$I$/d$V$ measurements, hybridization results in splitting and broadening of the observed peaks.[44,57] Further broadening and splitting can arise from intermolecular interactions within molecular assemblies.[44,58] In some cases, this broadening may even lead to $E_F$ crossing and partial charge transfer between the substrate and the molecular orbitals.[52,61,62] The additional charge density localized in the molecule arising from the charge transfer from/to the substrate is further stabilized by image charge screening in the metal substrate, leading to an additional downward shift of the molecular levels ($\Delta E_{pol.(CT)}$ in Fig. 6). The downward shift is counterbalanced by an increase in the local work function induced by the interfacial dipole associated with charge transfer ($\Delta E_{\mu(CT)}$).[52] This effect shifts the vacuum level upward relative to the molecular states, contributing to orbital pinning near the $E_F$.[52,62] In the case of Fermi level pinning, energy shifts caused by screening of the transferred charge and the local work function can therefore not be disentangled. Ultimately, the observed orbital alignment in d*I*/d*V* spectra reflects an equilibrium between image charge screening, hybridization, push-back effects, and dipole-induced vacuum level shifts.[52] In contrast, equation 1 is based on the assumption that the d*I*/d*V* curves acquired by tunnelling spectroscopy directly reflect the local density of states (LDOS). The obtained degree of charge transfer ($n_{e^-}$) should therefore not be interpreted quantitatively, but rather serves as an estimate for relative comparison between different $C_{60}$ species.

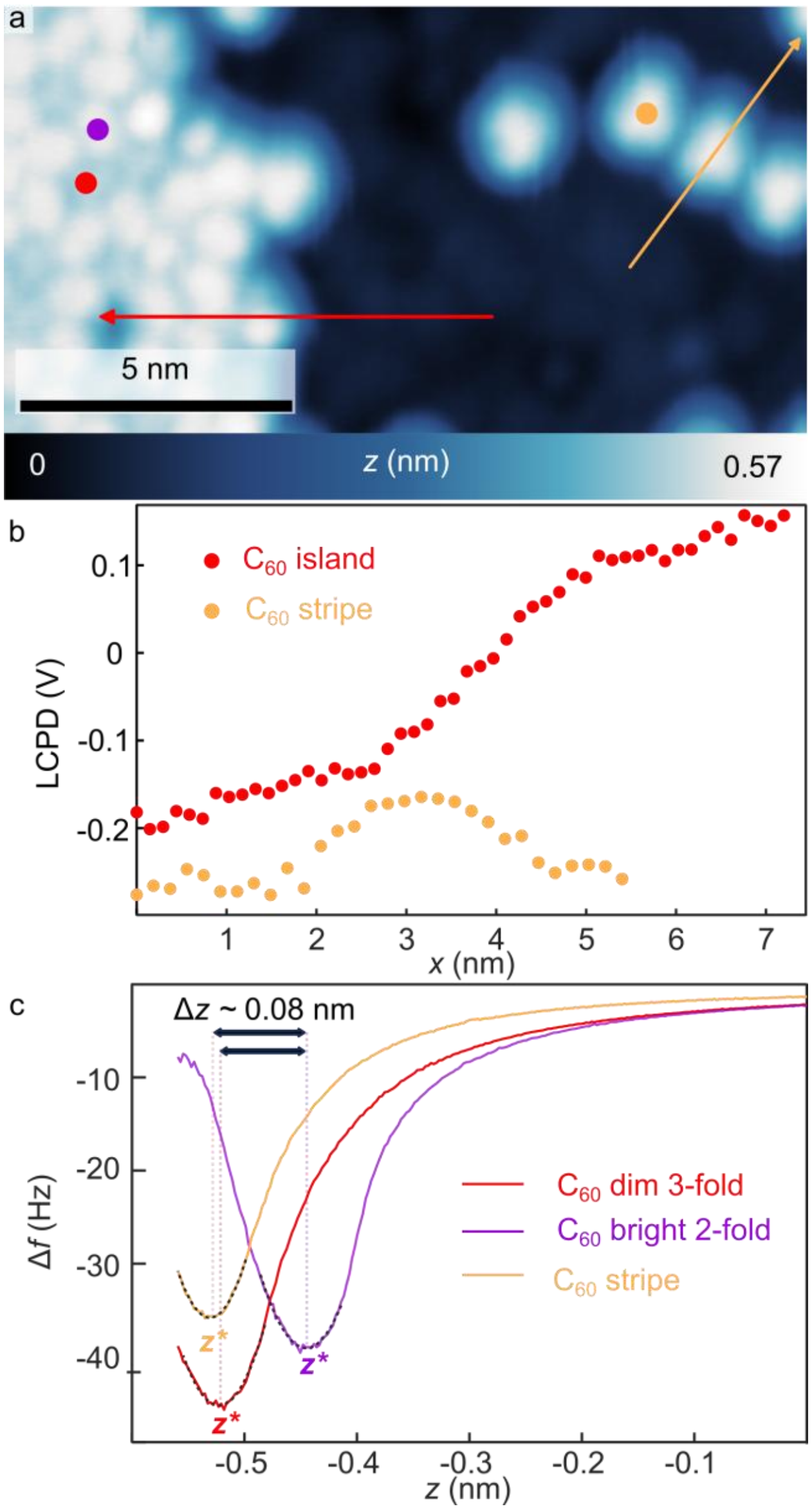

**Fig. 4.** KPFS and $\Delta f(z)$ characterization of the $C_{60}$ species. (a) Constant-current STM image of a $C_{60}$ island and the intermixed $C_{60}$/Zn-TAPP phase. (b) LCPD acquired at constant height along the paths indicated by the arrows in (a): across the edge of a $C_{60}$ island (red) and a $C_{60}$ stripe (orange). (c) $\Delta f(z)$ spectra acquired on different $C_{60}$ species (positions marked in (a)). Measurement parameters: (a) $I$ = 10 pA, $V$ = 0.25 V. (b) Feedback opened at $I$ = 10 pA, $V$ = 0.25 V, $z$ = 0.0 nm, $A$ = 0.1 nm. (c) Feedback opened at $I$ = 10 pA, $V$ = 0.25 V, $z$ = 0.0 nm; $\Delta f(z)$ spectra acquired at $V$ = 0 V, $A$ = 0.1 nm.

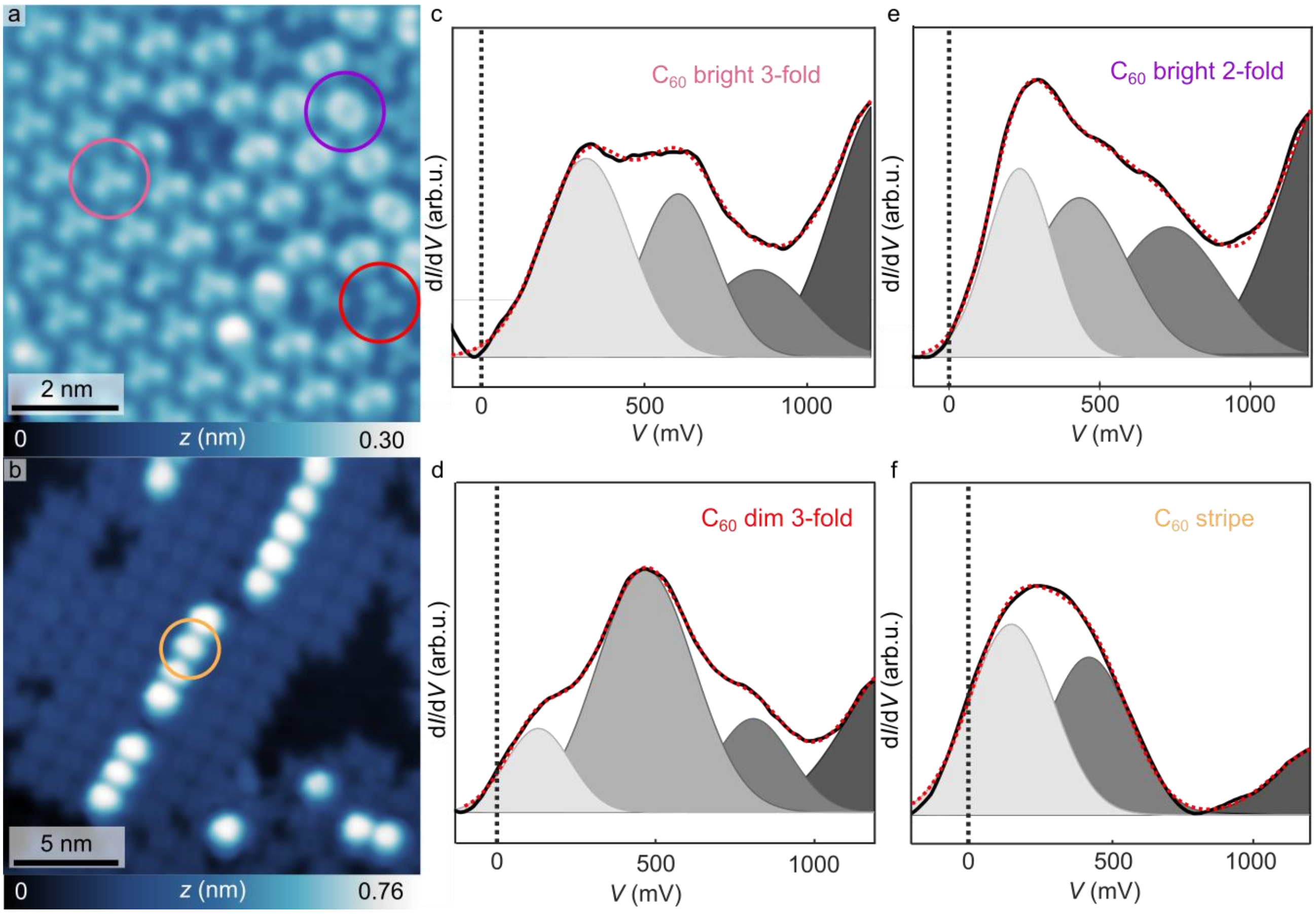

**Fig. 5.** d*I*/d*V* spectroscopy of the $C_{60}$ species. (a) Constant-current STM image of different adsorption geometries within a $C_{60}$ island. $C_{60}$ bright 3-fold, $C_{60}$ dim 3-fold (red) and $C_{60}$ bright 2-fold are highlighted in pink, red and purple, respectively. (b) Constant-current STM image of the intermixed $C_{60}$/Zn-TAPP phase. $C_{60}$ stripe is highlighted in orange. (c-e) d*I*/d*V* spectra acquired on $C_{60}$ bright 3-fold (c), $C_{60}$ dim 3-fold (d) and $C_{60}$ bright 2-fold (e). (f) d*I*/d*V* spectrum acquired on $C_{60}$ in the intermixed phase. Measurement parameters: (a) $I$ = 10 pA, $V$ = 0.2 V. (b) $I$ = 10 pA, $V$ = 0.6 V. (c-e) Feedback opened at $I$ = 19 pA, $V$ = 1.2 V. (f) Feedback opened at $I$ = 10 pA, $V$ = 1.2 V. The curves in (c-f) are fitted with 3 Gaussian for the $C_{60}$ LUMO and an additional Gaussian for the LUMO+1 (parameters are given in table S1). Areas of the Gaussians are indicated in grey scale and the resulting fits with dotted red lines. (c-f) Sample bias voltage modulation $V_{mod}$ = 0.04 $V_{rms}$.

## Discussion

For physisorbed molecules the image charge screening ($\Delta E_{pol.(sub.)}$) contribution has been estimated using a point charge at a distance *d* from the image plane (equation 2):[57,63]

$$\Delta E_{pol.(sub.)} = \frac{e^2}{4d} \qquad (2)$$

However, while this approach has been successfully applied to planar molecules,[63–65] it does not accurately capture the magnitude of substrate screening for $C_{60}$. In the latter case, the additional charge is partially delocalized over the spherical molecular geometry and likely preferentially distributed toward the metal-facing side rather than localized at a single point.[47] Consequently, a simple point-image model assuming a point charge at the molecular centre provides only a crude approximation of the actual screening behaviour (see Fig. S13).

For the present comparison, however, this uncertainty can be partially circumvented: dim 3-fold $C_{60}$ in the islands and $C_{60}$ in the one-dimensional rows adsorb at roughly the same height (see Fig. 4c) and show a 3-fold symmetry in STM (see Fig. 5a and Fig. S10), suggesting similar adsorption geometries. Therefore, the substrate screening term ($\Delta E_{pol.(sub.)}$) can be approximated to be effectively the same for both species and cancels in a differential analysis. The push-back effect can be treated accordingly, as it also depends on the adsorption height.[66] This approach allows to focus on relative energy shifts between the two environments, where the remaining differences can be attributed primarily to intermolecular (environmental) screening, as well as variations in the hybridization with neighbouring molecules.[57,63] Intermolecular interactions are stronger for $C_{60}$ in the close-packed islands than in the intermixed phase, resulting in a larger peak splitting (see Fig. 5) and being attributed to π-π orbital hybridization in the $C_{60}$ islands.[57]

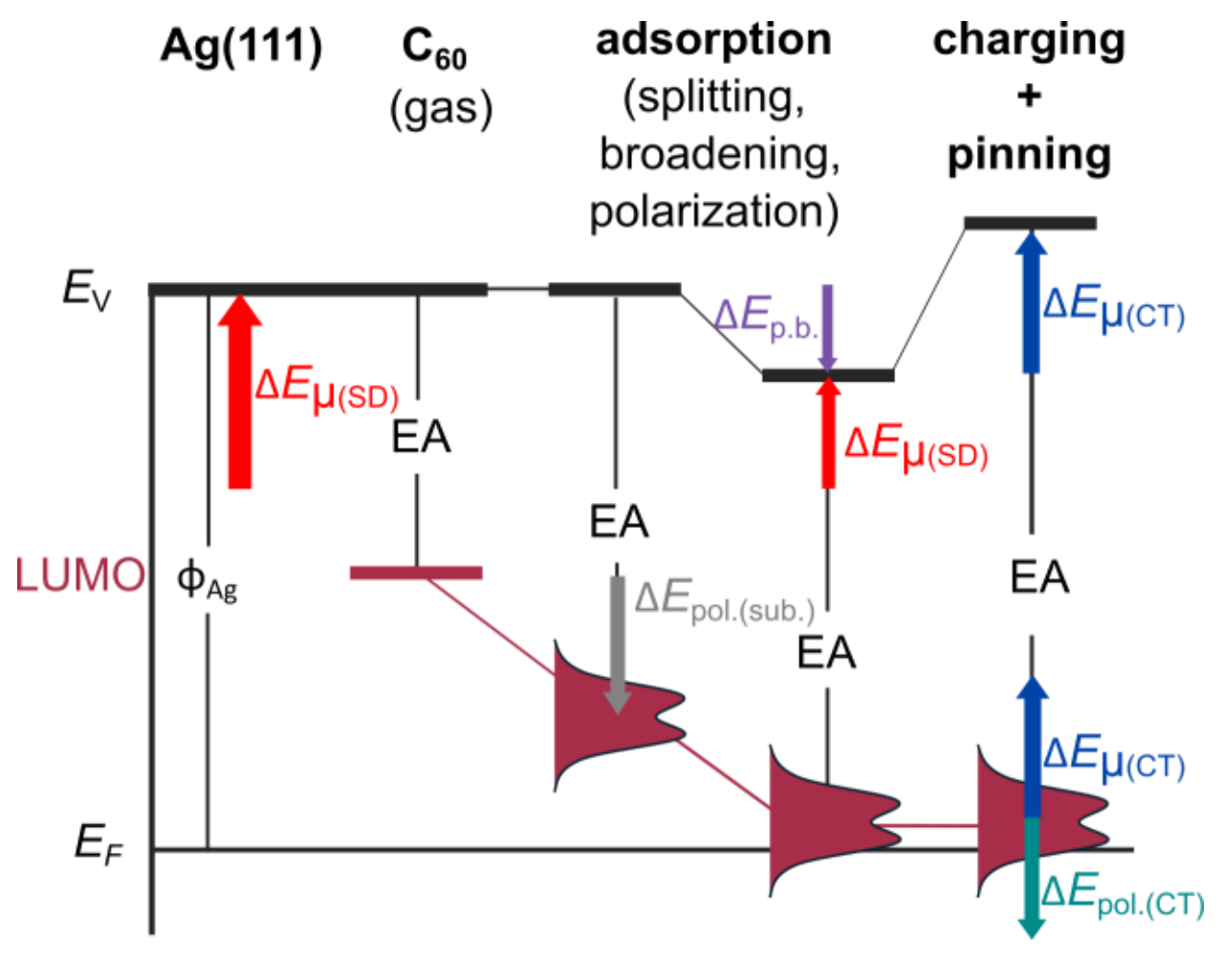

**Fig. 6.** Energy alignment of the $C_{60}$ LUMOs upon adsorption on Ag(111). Bold black lines show the relative position of the vacuum level influenced by changes in the local surface dipole (red and blue arrows). Alignment and peak shape of the $C_{60}$ LUMOs is shown in dark red. The contributions of the push-back effect (($\Delta E_{p.b.}$), the polarization by the substrate ($\Delta E_{pol.(sub.)}$), the surface dipole ($\Delta E_{\mu(SD)}$), the additional screening of the transferred charge ($\Delta E_{pol.(CT)}$) and the induced electrical dipole ($\Delta E_{\mu(CT)}$) are indicated with arrows.

According to Torrente et al.,[57] environmental screening ($\Delta E_{env.}$) depends mainly on the number $N$ of surrounding molecules, their polarizability $\alpha$, and the intermolecular distances $R$. Here, we considered $\Delta E_{env.}$ only for adjacent molecules (first-nearest neighbours), due to the $R^{-4}$ decay in equation 3:

$$\Delta E_{env.} = N \times \frac{e^2 \alpha}{2 \times R^4} \tag{3}$$

The gas-phase polarizabilities ($\boldsymbol{\alpha_0}$) for $C_{60}$ and Zn-TAPP were obtained from DFT (see SI). When using these gas-phase polarizabilities, for a $C_{60}$ molecule ($\boldsymbol{\alpha_0}$ = 549 a.u.) with six $C_{60}$ at $R$ = 1.00 nm (see Fig. S12b), equation 3 yields $\Delta E_{env.}$ = -351 meV. In the intermixed stripe phase, considering two $C_{60}$ and four Zn-TAPP ($\boldsymbol{\alpha_0}$ = 1024 a.u) and separations of 1.15/1.38/1.47 nm (see Fig. S12c), $\Delta E_{env}$ is about -228 meV. As a result, despite the larger Zn-TAPP polarizability, increased molecular separations result in a lower net environmental screening ($\Delta E_{env.}$).

The presence of the metallic substrate necessitates correcting the molecular polarizability $\alpha$ for image-charge screening according to Eq. 4 (see SI, Table S3):[67]

$$\alpha = \frac{\alpha_0}{\left(1 - \frac{\alpha_0}{4\pi\varepsilon_0 \times (2d)^3}\right)} \tag{4}$$

However, this correction requires the use of the simple point-image model resulting in complications for spherical molecules with small adsorption heights (see SI, Fig. S13). Even when locating the image charge at the centre of the molecule ($d$= 0.55 nm) – likely underestimating the screening of $C_{60}$ compared to Zn-TAPP – the difference in $\Delta E_{env.}$ between close-packed islands and one-dimensional stripe is minimal (≈ 40 meV) and cannot account for the observed shift in the d$I$/d$V$ spectra (Fig. S14).

After ruling out image-charge screening, push-back effect and estimating minimal differences in environmental screening, we attribute the observed shift of the LUMOs in the d$I$/d$V$ spectra (Fig. 5) to additional charging of $C_{60}$ in the intermixed stripe phase compared to dim 3-fold $C_{60}$ molecules. In this context, intermolecular charge transfer from Zn-TAPP to $C_{60}$ is unlikely, as there is no Zn-TAPP-related occupied orbital sufficiently close in energy to contribute (Fig. S11). Combined with the larger intermolecular distances in the intermixed phase this level alignment also rules out any effects of intermolecular hybridization.[57] Therefore, we ascribe the excess negative charge accumulation in the $C_{60}$ to increased substrate–molecule charge transfer in the one-dimensional stripes embedded within the otherwise neutral porphyrin scaffold.

DFT calculations for $C_{60}$ monomers and dimers (Fig. S15) support this assignment: charge analysis indicates that $C_{60}$ adsorbed at Ag atom vacancies carries more negative charge than on unreconstructed Ag(111) by about 0.1 $e$ (see SI Table S4). For monomers, the partial charge increases from ≈ 0.6 $e$ on the unreconstructed surface to ≈ 0.7 $e$ at a vacancy. Accordingly, $C_{60}$ adsorption energies increase by 0.1–0.3 eV per $C_{60}$ molecule when Ag atom vacancies are introduced.

In close-packed islands, adsorption on reconstructed sites seems to exclusively occur for dim 3-fold $C_{60}$ molecules (resulting in the dim contrast, shown for instance in Fig. 1b and 5a). Since our DFT calculations suggest lower charge transfer for unreconstructed sites, we attribute an even lower degree of charge transfer for the remaining (bright) species in the close-packed islands. Deconvolution of the LCPD (see Fig. 4b and SI) further corroborates enhanced local charge accumulation on $C_{60}$ within the one-dimensional stripe phase. Together, the tunnelling and force spectroscopy analysis demonstrates that $C_{60}$ in the intermixed phase attains a higher negative charge than $C_{60}$ in the close-packed islands.

These findings suggest an electrostatically governed mechanism for SAM formation. In close-packed islands, additional charging of $C_{60}$ is disfavoured by long-range Coulomb repulsion among neighbouring charged molecules. In contrast, in the one-dimensional intermixed phase, larger intermolecular separations and a lower $C_{60}$ coordination number reduce this repulsion, permitting greater charge uptake from the substrate. Therefore, the emergence of one-dimensional $C_{60}$ stripes can thus be understood as a balance between short-range attractive van der Waals dispersion forces, which promote molecular aggregation, and long-range Coulomb repulsion, which inhibits high local charge accumulation. Intercalation into the Zn-TAPP matrix spatially separates the $C_{60}$ molecules, thereby stabilizing a configuration in which $C_{60}$ exhibits a higher degree of charge transfer within the host framework.

## Conclusions

Using $C_{60}$ as a model acceptor, we have shown how electrostatic interactions govern both the assembly and electronic structure of SAMs on Ag(111). KPFS indicates substrate-mediated charging of $C_{60}$, and, after accounting for spatial averaging, reveals enhanced charge accumulation in one-dimensional stripe phases relative to two-dimensional close-packed islands. Consistently, d*I*/d*V* spectroscopy confirms a larger partial occupation of the $C_{60}$ LUMOs in the stripe phase. We attribute this effect to the different lateral electrostatic environments in the two phases. In close-packed islands, strong lateral repulsion raises the chemical potential for additional electron uptake and limits LUMO filling. In the one-dimensional intermixed stripes, increased intermolecular spacing and fewer $C_{60}$ neighbours reduce the electrostatic penalty, enabling higher charging. These results establish molecular packing density and dimensionality as effective levers for tuning charge transfer at metal–molecule interfaces.

This approach suggests that rational variation of the template structure could be used to tune $C_{60}$ charging by systematically modifying the local electrostatic environment in which it is embedded. Extending this concept to templates with different geometries, periodicities, and intermolecular interaction strengths could therefore provide a route to spatially and hierarchically control both $C_{60}$ organization and its degree of charging, offering a general strategy for designing molecular assemblies with coupled structural and electronic functionality.

## Experimental

The Ag(111) single crystal was supplied by SPL and cleaned using alternating cycles of $Ne^+$ sputtering (1 keV, 15 min, $p = 5*10^{-6}$ mbar) and subsequent annealing to 667 K. Zn(II)-5,10,15,20-Tetrakis-(4-aminophenyl)-porphyrin (Zn-TAPP, 98%) was purchased from Porphychem and $C_{60}$ (99.5%) was ordered from Sigma Aldrich. Both molecules were evaporated with a 3-cell Knudsen cell evaporator (Kentax GmbH). Before evaporation on the sample the evaporation rate was determined using a quartz micro balance (QMB). Sub monolayer coverage was achieved by evaporating ~5 Hz of both molecules onto the Ag(111) single crystal kept at 300 K. All LT-SPM measurements were conducted at 77 K and ultra-high vacuum (~$5*10^{-11}$mbar) in a CreaTec Fischer GmbH LT-SPM instrument (bath cryostat with liquid $N_2$). Bias voltage ($V$) is given as sample bias with respect to the tip. The surface was analyzed with STM using a Pt/Ir tip or a qPlus sensor with an etched Pt/Ir tip. For the AFM measurements oscillation amplitudes in the range of 0.1-0.3 nm were used. Measurements in the Lock-in mode were conducted by adding a sinusoidal voltage modulation ($V_{mod}$) at 724 Hz to the bias voltage ($V$).

Models of the Zn-TAPP and $C_{60}$ molecules as the ones shown in Figures 1b,c,f were built in Avogadro (an open-source molecular builder and visualization tool. Version 1.2.0n http://avogadro.cc/).[68]

DFT calculations were performed in VASP, version 5.4.4.[69,70] As a model, the Ag(111) surface with 254 Ag atoms (four layers) and 320 Ag atoms (five layers) with the initial separation of 2.889 Å was employed along with D2 dispersion correction, PBE functional and the projector-augmented wave (PAW) method.[71] In the *z* direction, the cell length was set to 40 Å. Only the gamma point was used in the reciprocal space. First, all models were optimized with the energy cut-off of 250 eV and four layers, bottom two of which were held frozen during optimization to mimic bulk effects. The fullerene molecules were adsorbed in various configurations and only the most stable ones were considered further. The obtained structures were re-optimized with an additional layer of frozen Ag atoms at the cut-off of 300 eV. Final single-point energy evaluations were performed at the energy cut-off of 600 eV. Charges on atoms were analysed using the Bader approach.[72]

## Conflicts of interest

There are no conflicts to declare.

## Data availability

The data supporting this article have been included as part of the supplementary information (SI). Supplementary information: Fig. S1-15, Table S1-4

## Acknowledgements

B.A., M.T., M.Z. and L.L.P. acknowledge support from the ERC under the European Union's Horizon 2020 research and innovation programme (grant agreement no. 101039746) funded by the European Union. The computational results presented have been in part achieved using the HPC infrastructure LEO of the University of Innsbruck.

## References

1 M. Bouachrine and S. Ayachi, *RSC Adv.*, 2025, **15**, 45177–45195.

2 N. Cernetic, T. Weidner, J. E. Baio, H. Lu, H. Ma and A. K.-Y. Jen, *Advanced Functional Materials*, 2015, **25**, 5376–5383.
3 H. Bronstein, C. B. Nielsen, B. C. Schroeder and I. McCulloch, *Nat Rev Chem*, 2020, **4**, 66–77.
4 C. Kunkel, J. T. Margraf, K. Chen, H. Oberhofer and K. Reuter, *Nat Commun*, 2021, **12**, 2422.
5 A. M. Nardes, S. Ahn, D. Rourke, C. Mao, J. van de Lagemaat, A. J. Ferguson, W. Park and N. Kopidakis, *Organic Electronics*, 2016, **39**, 59–63.
6 T. Aihara, S. A. Abd-Rahman and H. Yoshida, *Phys. Rev. B*, 2021, **104**, 085305.
7 S. Casalini, C. Augusto Bortolotti, F. Leonardi and F. Biscarini, *Chemical Society Reviews*, 2017, **46**, 40–71.
8 B. H. Hamadani, D. A. Corley, J. W. Ciszek, J. M. Tour and D. Natelson, *Nano Lett.*, 2006, **6**, 1303–1306.
9 J. Kim, Y. S. Rim, Y. Liu, A. C. Serino, J. C. Thomas, H. Chen, Y. Yang and P. S. Weiss, *Nano Lett.*, 2014, **14**, 2946–2951.
10 K.-Y. Wu, S.-Y. Yu and Y.-T. Tao, *Langmuir*, 2009, **25**, 6232–6238.
11 Y. Xu, K.-J. Baeg, W.-T. Park, A. Cho, E.-Y. Choi and Y.-Y. Noh, *ACS Appl. Mater. Interfaces*, 2014, **6**, 14493–14499.
12 R. Ruvalcaba, Z. Liu, M. I. Nugraha, G. Harrison, Y.-Y. Yang, M. Thaler, A. V. Marsh, M. Zeilerbauer, C. Aivalioti, R. R. Aguileta-Vazquez, L. Tsetseris, L. L. Patera, P. Zahl, M. Heeney, T. D. Anthopoulos and S. Fatayer, *ACS Materials Lett.*, 2025, **7**, 1421–1430.
13 I. Hehn, S. Schuster, T. Wächter, T. Abu-Husein, A. Terfort, M. Zharnikov and E. Zojer, *J Phys Chem Lett*, 2016, **7**, 2994–3000.
14 J. L. Cabellos, D. J. Mowbray, E. Goiri, A. El-Sayed, L. Floreano, D. G. de Oteyza, C. Rogero, J. E. Ortega and A. Rubio, *J. Phys. Chem. C*, 2012, **116**, 17991–18001.
15 J. W. Pinder, G. H. Major, D. R. Baer, J. Terry, J. E. Whitten, J. Čechal, J. D. Crossman, A. J. Lizarbe, S. Jafari, C. D. Easton, J. Baltrusaitis, M. A. van Spronsen and M. R. Linford, *Applied Surface Science Advances*, 2024, **19**, 100534.
16 G. Binnig, H. Rohrer, Ch. Gerber and E. Weibel, *Phys. Rev. Lett.*, 1982, **49**, 57–61.
17 G. Binnig, C. F. Quate and Ch. Gerber, *Phys. Rev. Lett.*, 1986, **56**, 930–933.
18 A. P. Labonté, S. L. Tripp, R. Reifenberger and A. Wei, *J. Phys. Chem. B*, 2002, **106**, 8721–8725.
19 G. Yang and G. Liu, *J. Phys. Chem. B*, 2003, **107**, 8746–8759.
20 Q. Guo and F. Li, *Phys. Chem. Chem. Phys.*, 2014, **16**, 19074.
21 L. Gross, F. Mohn, N. Moll, P. Liljeroth and G. Meyer, *Science*, 2009, **325**, 1110–1114.
22 F. Albrecht, N. Pavliček, C. Herranz-Lancho, M. Ruben and J. Repp, *J. Am. Chem. Soc.*, 2015, **137**, 7424–7428.
23 B. Schuler, W. Liu, A. Tkatchenko, N. Moll, G. Meyer, A. Mistry, D. Fox and L. Gross, *Phys. Rev. Lett.*, 2013, **111**, 106103.
24 F. Albrecht, F. Bischoff, W. Auwärter, J. V. Barth and J. Repp, *Nano Lett.*, 2016, **16**, 7703–7709.
25 L. Gross, F. Mohn, P. Liljeroth, J. Repp, F. J. Giessibl and G. Meyer, *Science*, 2009, **324**, 1428–1431.
26 L. Gross, B. Schuler, F. Mohn, N. Moll, N. Pavliček, W. Steurer, I. Scivetti, K. Kotsis, M. Persson and G. Meyer, *Phys. Rev. B*, 2014, **90**, 155455.
27 S. Limbu, K.-B. Park, J. Wu, H. Cha, S. Yun, S.-J. Lim, H. Yan, J. Luke, G. Ryu, C.-J. Heo, S. Kim, Y. W. Jin, J. R. Durrant and J.-S. Kim, *ACS Nano*, 2021, **15**, 1217–1228.
28 F. Castet, G. D'Avino, L. Muccioli, J. Cornil and D. Beljonne, *Phys. Chem. Chem. Phys.*, 2014, **16**, 20279–20290.
29 M. S. Ryu and J. Jang, *Solar Energy Materials and Solar Cells*, 2010, **94**, 1384–1388.
30 T. D. M. Bell, T. A. Smith, K. P. Ghiggino, M. G. Ranasinghe, M. J. Shephard and M. N. Paddon-Row, *Chemical Physics Letters*, 1997, **268**, 223–228.
31 D. M. Guldi, *Chem. Soc. Rev.*, 2002, **31**, 22–36.
32 F. Spänig, M. Ruppert, J. Dannhäuser, A. Hirsch and D. M. Guldi, *J. Am. Chem. Soc.*, 2009, **131**, 9378–9388.
33 M. Di Marino, F. Sedona, M. Sambi, T. Carofiglio, E. Lubian, M. Casarin and E. Tondello, *Langmuir*, 2010, **26**, 2466–2472.
34 F. Sedona, M. Di Marino, A. Basagni, L. Colazzo and M. Sambi, *J. Phys. Chem. C*, 2014, **118**, 1587–1593.
35 W. Jin, Q. Liu, D. B. Dougherty, W. G. Cullen, J. E. Reutt-Robey, J. Weeks and S. W. Robey, *The Journal of Chemical Physics*, 2015, **142**, 101910.
36 G. Tomba, M. Stengel, W.-D. Schneider, A. Baldereschi and A. De Vita, *ACS Nano*, 2010, **4**, 7545–7551.
37 B. Schuler, S.-X. Liu, Y. Geng, S. Decurtins, G. Meyer and L. Gross, *Nano Lett.*, 2014, **14**, 3342–3346.
38 K. Biswas, J. I. Urgel, K. Xu, J. Ma, A. Sánchez-Grande, P. Mutombo, A. Gallardo, K. Lauwaet, B. Mallada, B. de la Torre, A. Matěj, J. M. Gallego, R. Miranda, P. Jelínek, X. Feng and D. Écija, *Angewandte Chemie International Edition*, 2021, **60**, 25551–25556.
39 W. W. Pai, H. T. Jeng, C.-M. Cheng, C.-H. Lin, X. Xiao, A. Zhao, X. Zhang, G. Xu, X. Q. Shi, M. A. Van Hove, C.-S. Hsue and K.-D. Tsuei, *Phys. Rev. Lett.*, 2010, **104**, 036103.
40 A. D. Pia, M. Riello, D. Stassen, T. S. Jones, D. Bonifazi, A. D. Vita and G. Costantini, *Nanoscale*, 2016, **8**, 19004–19013.
41 C. D. Reddy, Z. Gen Yu and Y.-W. Zhang, *Sci Rep*, 2015, **5**, 12221.
42 H. I. Li, K. Pussi, K. J. Hanna, L.-L. Wang, D. D. Johnson, H.-P. Cheng, H. Shin, S. Curtarolo, W. Moritz, J. A. Smerdon, R. McGrath and R. D. Diehl, *Phys. Rev. Lett.*, 2009, **103**, 056101.
43 K. Pussi, H. I. Li, H. Shin, L. N. Serkovic Loli, A. K. Shukla, J. Ledieu, V. Fournée, L. L. Wang, S. Y. Su, K. E. Marino, M. V. Snyder and R. D. Diehl, *Phys. Rev. B*, 2012, **86**, 205406.
44 X.-Q. Shi, M. A. Van Hove and R.-Q. Zhang, *J Mater Sci*, 2012, **47**, 7341–7355.
45 H. Shin, A. Schwarze, R. D. Diehl, K. Pussi, A. Colombier, É. Gaudry, J. Ledieu, G. M. McGuirk, L. N. Serkovic Loli, V. Fournée, L. L. Wang, G. Schull and R. Berndt, *Phys. Rev. B*, 2014, **89**, 245428.
46 T. Chutora, J. Redondo, B. De La Torre, M. Švec, P. Jelínek and H. Vázquez, *Beilstein J. Nanotechnol.*, 2017, **8**, 1073–1079.
47 X. Lu, M. Grobis, K. H. Khoo, S. G. Louie and M. F. Crommie, *Phys. Rev. B*, 2004, **70**, 115418.
48 J. A. Larsson, S. D. Elliott, J. C. Greer, J. Repp, G. Meyer and R. Allenspach, *Phys. Rev. B*, 2008, **77**, 115434.
49 L.-L. Wang and H.-P. Cheng, *Phys. Rev. B*, 2004, **69**, 165417.
50 F. Mohn, L. Gross, N. Moll and G. Meyer, *Nature Nanotech*, 2012, **7**, 227–231.
51 S. Trishin, T. Müller, D. Rolf, C. Lotze, P. Rietsch, S. Eigler, B. Meyer and K. J. Franke, *J. Phys. Chem. C*, 2022, **126**, 7667–7673.

52 M. Hollerer, D. Lüftner, P. Hurdax, T. Ules, S. Soubatch, F. S. Tautz, G. Koller, P. Puschnig, M. Sterrer and M. G. Ramsey, *ACS Nano*, 2017, **11**, 6252–6260.
53 M. Bommert, A. Kinikar, M. A. Moro, N. Merino Diez, F. Xiang, N. Krane, N. Bassi, R. Widmer and O. Gröning, *Carbon*, 2024, **228**, 119370.
54 S. Trishin, T. Müller, D. Rolf, C. Lotze, P. Rietsch, S. Eigler, B. Meyer and K. J. Franke, *J. Phys. Chem. C*, 2022, **126**, 7667–7673.
55 U. Zerweck, C. Loppacher, T. Otto, S. Grafström and L. M. Eng, *Nanotechnology*, 2007, **18**, 084006.
56 G. Elias, T. Glatzel, E. Meyer, A. Schwarzman, A. Boag and Y. Rosenwaks, *Beilstein J. Nanotechnol.*, 2011, **2**, 252–260.
57 I. Fernández Torrente, K. J. Franke and J. Ignacio Pascual, *J. Phys.: Condens. Matter*, 2008, **20**, 184001.
58 N. Haag, D. Lüftner, F. Haag, J. Seidel, L. L. Kelly, G. Zamborlini, M. Jugovac, V. Feyer, M. Aeschlimann, P. Puschnig, M. Cinchetti and B. Stadtmüller, *Phys. Rev. B*, 2020, **101**, 165422.
59 G. Zamborlini, D. Lüftner, Z. Feng, B. Kollmann, P. Puschnig, C. Dri, M. Panighel, G. Di Santo, A. Goldoni, G. Comelli, M. Jugovac, V. Feyer and C. M. Schneider, *Nat Commun*, 2017, **8**, 335.
60 R. Otero, A. L. Vázquez De Parga and J. M. Gallego, *Surface Science Reports*, 2017, **72**, 105–145.
61 O. T. Hofmann, P. Rinke, M. Scheffler and G. Heimel, *ACS Nano*, 2015, **9**, 5391–5404.
62 S. Braun and W. R. Salaneck, *Chemical Physics Letters*, 2007, **438**, 259–262.
63 K. A. Cochrane, A. Schiffrin, T. S. Roussy, M. Capsoni and S. A. Burke, *Nat Commun*, 2015, **6**, 8312.
64 D. Kumar, C. Krull, Y. Yin, N. V. Medhekar and A. Schiffrin, *ACS Nano*, 2019, **13**, 11882–11890.
65 D. Kumar, J. Hellerstedt, B. Lowe and A. Schiffrin, *Nanotechnology*, 2023, **34**, 205601.
66 P. C. Rusu, G. Giovannetti, C. Weijtens, R. Coehoorn and G. Brocks, *J. Phys. Chem. C*, 2009, **113**, 9974–9977.
67 C. M. Teodorescu, *Phys. Chem. Chem. Phys.*, 2015, **17**, 21302–21314.
68 M. D. Hanwell, D. E. Curtis, D. C. Lonie, T. Vandermeersch, E. Zurek and G. R. Hutchison, *J Cheminform*, 2012, **4**, 17.
69 G. Kresse and J. Furthmüller, *Phys. Rev. B*, 1996, **54**, 11169–11186.
70 G. Kresse and D. Joubert, *Phys. Rev. B*, 1999, **59**, 1758–1775.
71 P. E. Blöchl, *Phys. Rev. B*, 1994, **50**, 17953–17979.
72 W. Tang, E. Sanville and G. Henkelman, *J. Phys.: Condens. Matter*, 2009, **21**, 084204.

### SI references

73 F. Neese, *WIREs Comput. Mol. Sci.*, 2025, **15**, e70019.
74 S. Grimme, S. Ehrlich and L. Goerigk, *J. Comput. Chem.*, 2011, **32**, 1456–1465.
75 A. D. Becke, *J. Chem. Phys.*, 1993, **98**, 5648–5652.
76 C. Lee, W. Yang and R. G. Parr, *Phys. Rev. B*, 1988, **37**, 785–789.
77 F. Weigend and R. Ahlrichs, *Phys. Chem. Chem. Phys*., 2005, **7**, 3297–3305.
78 F. Weigend, *Phys. Chem. Chem. Phys.*, 2006, **8**, 1057.
79 K. Eichkorn, F. Weigend, O. Treutler and R. Ahlrichs, *Theor. Chem. Acc.*, 1997, **97**, 119–124.
80 S. Kossmann and F. Neese, *Chem. Phys. Lett.*, 2009, **481**, 240–243.
81 D. Rappoport and F. Furche, *J. Chem. Phys.*, 2010, **133**, 134105.
82 D. Rappoport, *J. Chem. Phys.*, 2021, **155**, 124102.
83 W. Hieringer, K. Flechtner, A. Kretschmann, K. Seufert, W. Auwärter, J. V. Barth, A. Görling, H.-P. Steinrück and J. M. Gottfried, *J. Am. Chem. Soc.*, 2011, **133**, 6206–6222.
84 P. Amsalem, J. Niederhausen, A. Wilke, G. Heimel, R. Schlesinger, S. Winkler, A. Vollmer, J. P. Rabe and N. Koch, *Phys. Rev. B*, 2013, **87**, 035440.

# Supporting information

## Steering interfacial molecular structures by substrate-molecule charge transfer

Benjamin Achatz,*[a] Matthias Zeilerbauer,[a] Marco Thaler,[a] Milan Ončák,[b] Shadi Fatayer,[c] and Laerte L. Patera*[a]

a. Department of Physical Chemistry, University of Innsbruck, 6020 Innsbruck (Austria)
E-mail: benjamin.achatz@uibk.ac.at, laerte.patera@uibk.ac.at

b. Department of Ion Physics and Applied Physics, University of Innsbruck, 6020 Innsbruck (Austria)

c. Physical Science and Engineering Division (PSE), King Abdullah University of Science and Technology (KAUST), Thuwal 23955-6900 (Saudi Arabia)

### Epitaxial relationship of the stripe phase

In Fig. S1 the lattice vectors of the Ag(111) surface have been obtained from the (2√3x√3)R30° superstructure of $C_{60}$.[44] Investigation of the epitaxial relationship revealed six distinct orientations of the $C_{60}$ stripes (Fig. S2). Importantly, the $C_{60}$ stripe directions always follow the surrounding porphyrin lattice, as already observed for the stripe phase on Ag(110).[34] In contrast, the stripes do not coincide with the lattice vectors of the (2√3x√3)R30° superstructure, suggesting that the molecular rearrangement upon intercalation of $C_{60}$ into the porphyrin lattice dictates the growth directions of the $C_{60}$ stripes, rather than epitaxial matching of $C_{60}$ with the underlying substrate. To further investigate these structural changes, we compared models of the unit cells of the porphyrin layer and the $C_{60}$ stripe phase extracted from Fig. 1e. The obtained structures can be related to commensurate models (Fig. S3) within an error of below 5% for all lattice parameters. The epitaxial relationships of the commensurate structures with respect to the basis vectors of the Ag(111) substrate are given by the matrices M in Fig. S3. The models shown in Fig. S3 reveal that all porphyrins are oriented along the [-101] direction of the Ag(111) substrate. $C_{60}$ intercalation requires an expansion of the porphyrin lattice which can be described as a translation of two porphyrins along the [0-11] direction, by twice the lattice constant of Ag(111) (red arrow in Fig. S3). Diffraction measurements would be required to confirm these commensurate structures, but they lie outside the scope of this study.

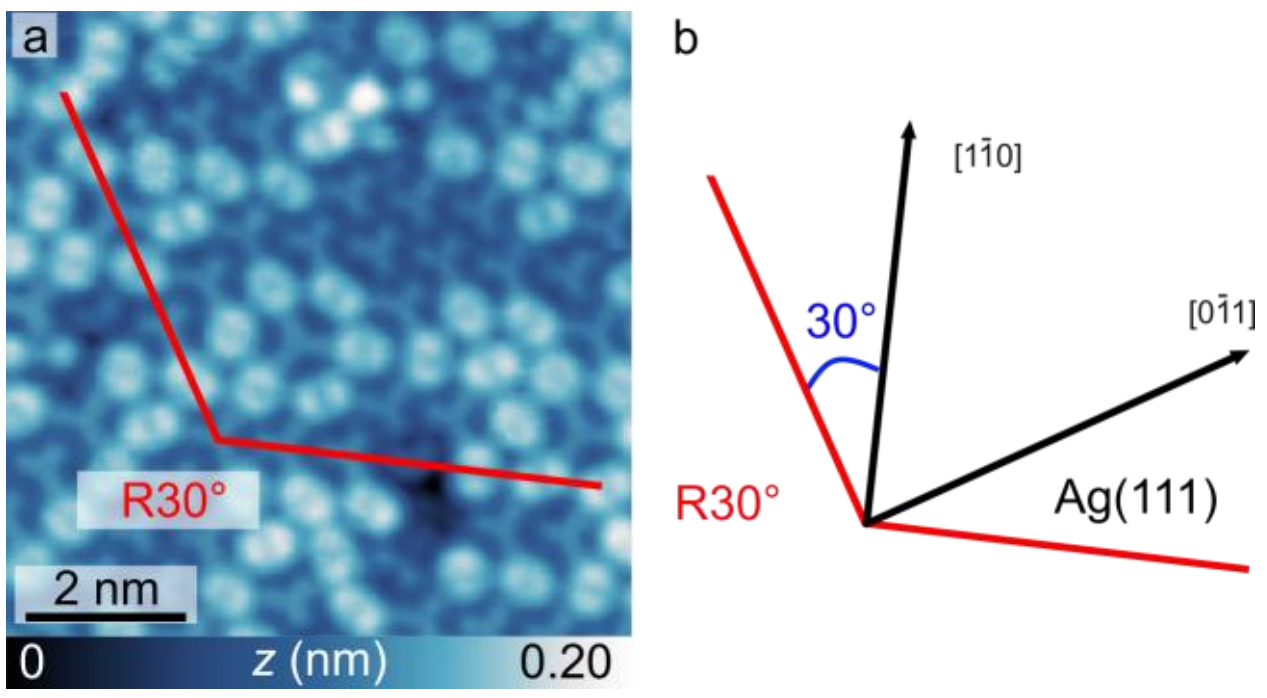

**Fig. S1.** (2√3x√3)R30° $C_{60}$ superstructure on Ag(111). (a) Constant-current STM image of a $C_{60}$ island on Ag(111) (lattice vectors indicated in red). (b) Inferred crystallographic directions on the Ag(111) surface (black). Measurement parameters: (a) $I$ = 10 pA, $V$ = 0.2 V.

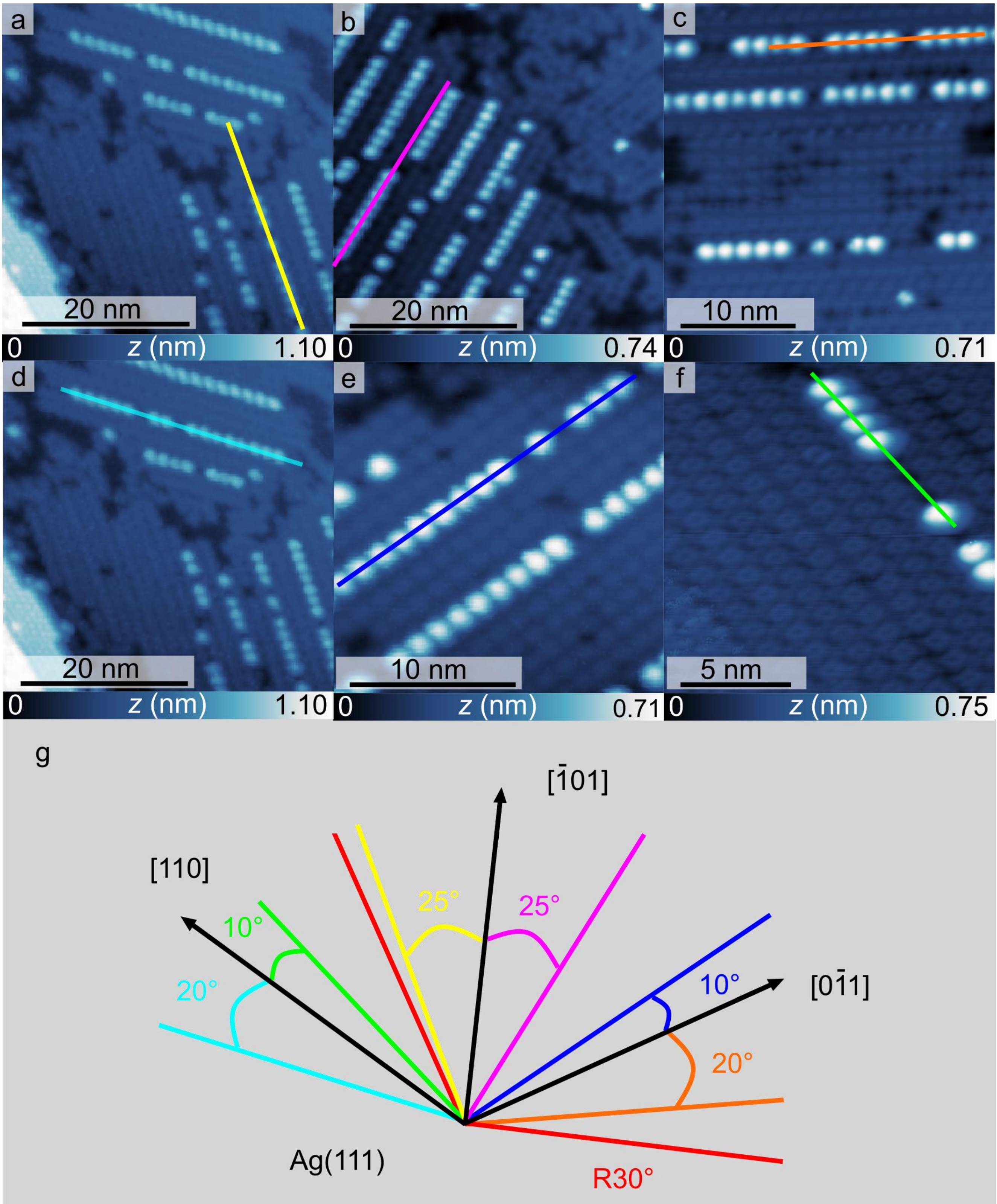

**Fig. S2.** $C_{60}$ stripes growth directions. (a-f) Constant-current STM image of $C_{60}$ chains on Ag(111) (g) Inferred growth direction of the $C_{60}$ chain in (a-f) compared to the lattice on the Ag(111) surface (black) and the (2√3x√3)R30° $C_{60}$ superstructure (red). Measurement parameters: (a,d) $I$ = 8 pA, $V$ = 1.0 V (b,e) $I$ = 5 pA, $V$ = 0.6 V (c) $I$ = 20 pA, $V$ = 0.5 V (f) $I$ = 6 pA, $V$ = 0.4 V.

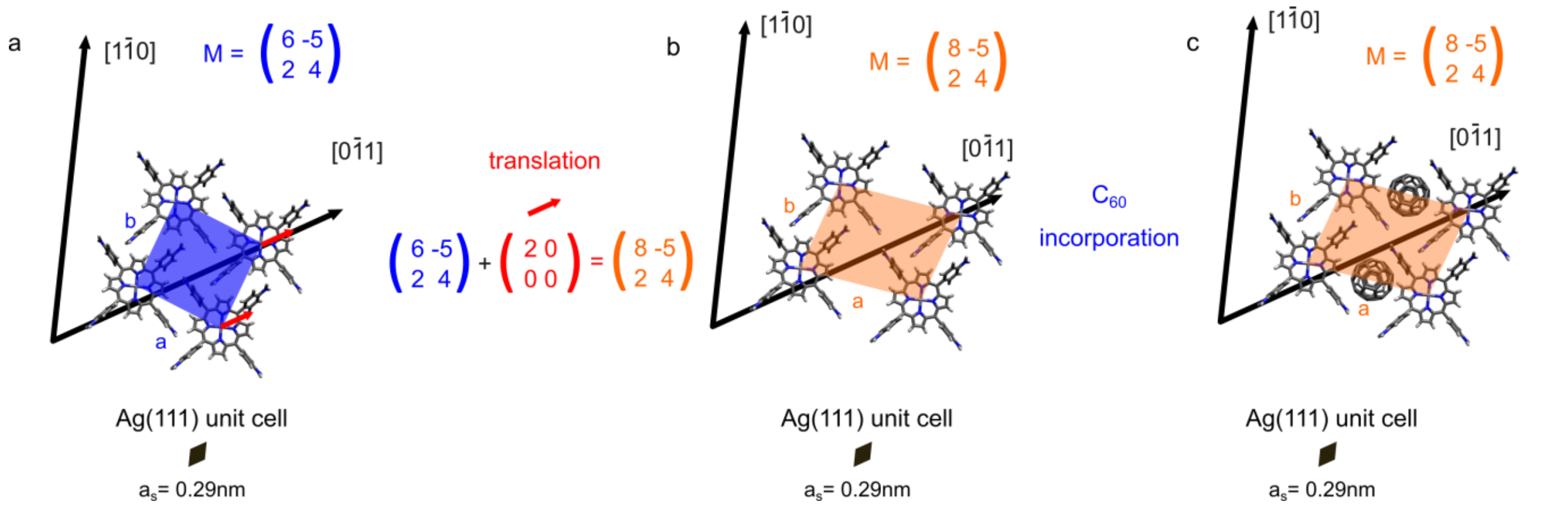

**Fig. S3.** Structural changes in the porphyrin lattice during $C_{60}$ intercalation for commensurate models. Commensurate model of (a) pristine Zn-TAPP lattice, (b) Zn-TAPP lattice after translation of two porphyrins and (c) Zn-TAPP lattice hosting one $C_{60}$ per unit cell. Epitaxial relationship to the Ag(111) basis vectors is given by the matrices M. Model based on overlays of Fig. 1e.

## KPFS

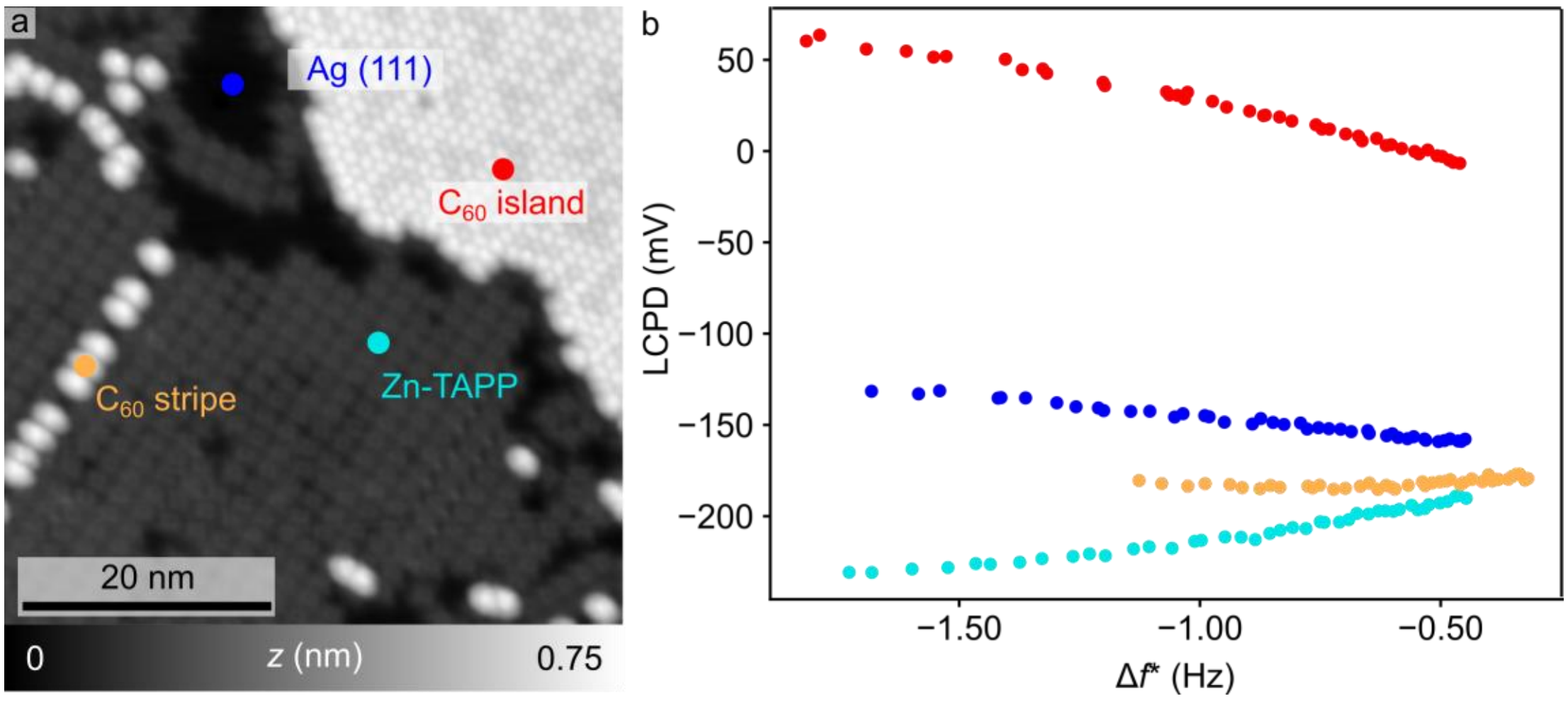

**Fig. S4.** Height-dependent KPFS. (a) Constant-current STM image acquired after annealing to 428 K for 3 h. (b) Fitted LCPD values acquired at the positions marked in (a) as a function of Δ*f**. Measurement parameters: (a) *I* = 20 pA, *V* = 0.5 V; (b) Feedback opened on the marked positions at *I* = 5 pA, *V* = 0.6 V, *A* = 0.3 nm.

## Lateral Resolution in KPFM

The lateral resolution of KPFM is limited due to the averaging of long-range electrostatic forces. To estimate the lateral resolution in our experiments, we acquired KPFS spectra at different tip-sample distances across the interface between $C_{60}$ and Zn-TAPP islands (marked by an arrow in Fig. S5a). Figure S5b shows the decreasing LCPD contrast between the $C_{60}$ island and adjacent Zn-TAPP layer with increasing tip height (*z*). The LCPD profiles were fitted with an error function (Fig. S5c) and the resolution was estimated by calculating the full width half maximum (FHWM) of the gaussian component from its standard deviation ($\sigma$).

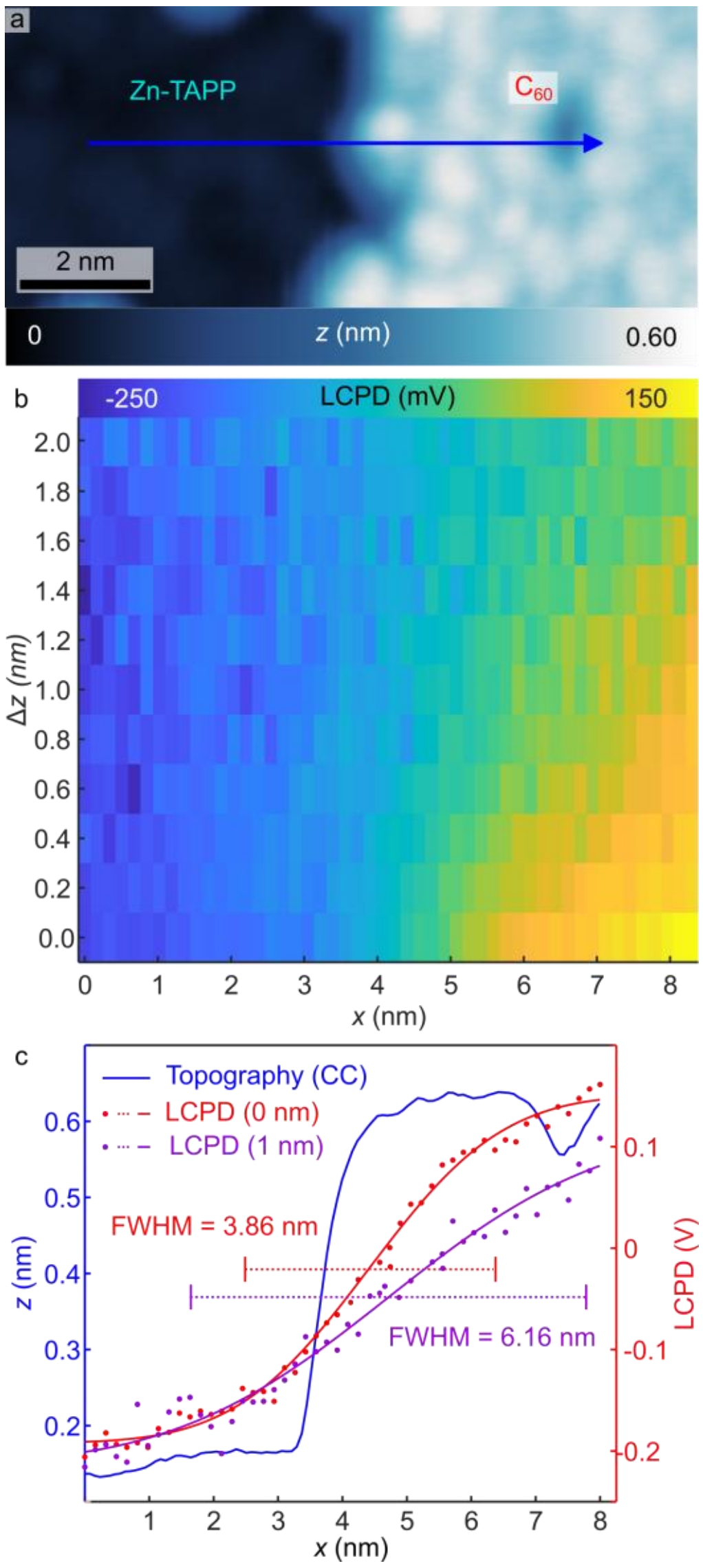

**Fig. S5.** Tip-height dependence of LCPD spatial resolution. (a) Constant-current STM image. (b) LCPD acquired across a step edge, indicated by the arrow in (a), at different tip-sample distances in increments of 0.2 nm. (c) STM topographic line profile extracted form image (a) and extracted LCPD at $z$ = 0 nm (FWHM = 3.86 nm $\sigma$ = 1.64 nm) and $z$ = 1 nm (FWHM = 6.16 nm and $\sigma$ = 2.61 nm). Measurement parameters: (a) tunneling current $I$ = 10 pA, sample bias voltage $V$ = 0.25 V. (b-c) Feedback opened at $I$ = 10 pA, $V$ = 0.25 V, $A$ = 0.3 nm.

To estimate the intrinsic LCPD of the $C_{60}$ stripe from the measured LCPD, we modelled the measurement as the convolution of the surface dipole variation with a point-spread function (PSF).[84] A gaussian PSF and a double step function were used for the deconvolution. This simplified modeling of the PSF and the intrinsic LCPD qualitatively describes the smeared out contrast in KPFS. The width of the double step function was set to 0.7 nm, which roughly corresponds to the size of a $C_{60}$ molecule. By fitting the experimental data shown in Fig. S6a, we determined a FWHM (1.85±0.15 nm) of the PSF ($FWHM_{PSF}$) – which is considerably smaller than the one determined for the gaussian component of the error function in Fig. S6 (FWHM = 3.86±0.18 nm). This value of the $FWHM_{PSF}$ can be then used to determine the total contribution to the measured LCPD of the $C_{60}$ molecules in a one-dimensional stripe phase. We modeled the $C_{60}$ stripe phase as seven two-dimensional disks with a diameter of 0.7 nm, arranged collinearly with a center-to-center spacing of 1.47 nm, to emulate the $C_{60}$ stripe-phase assembly (Fig. S6b). The object was then convolved with a Gaussian point-spread function (PSF) using the fitted $FWHM_{PSF}$. We then evaluated, at the position of the central disk, the contribution of each individual disk to the deconvolved signal (Fig. S3b). Only the center molecule and its two nearest neighbors contribute appreciably (see Fig. S6c), indicating that measured LCPD in Fig. 4b reflects just 12.8% of the intrinsic stripe contrast (yielding a renormalization factor of 7.8±0.6). An additional profile acquired at $z$ = 0.1 nm (Fig. S7) yielded a similar renormalization factor of 8.2±1.0. The increased error on the renormalization factor can be attributed to lower signal to noise ratio in KPFS upon retracting the tip. While these deconvolutions do not provide quantitative information about the charge in the $C_{60}$ stripes they can be used to compare the magnitude of the LCPD with the two-dimensional $C_{60}$ islands. In contrast to the stripe case, the $C_{60}$ islands have lateral dimensions that are much larger than the $FWHM_{PSF}$ (see Fig. S8a). Therefore, the contrast measured 5 nm inside the island is effectively insensitive to PSF convolution (Fig. S8b). According to this simplified model the deconvolved LCPD of the stripe phase is around eight times larger than the measured LCPD, supporting the assumption that averaging effects heavily suppresses the contrast for one-dimensional stripe phases, compared to the islands.

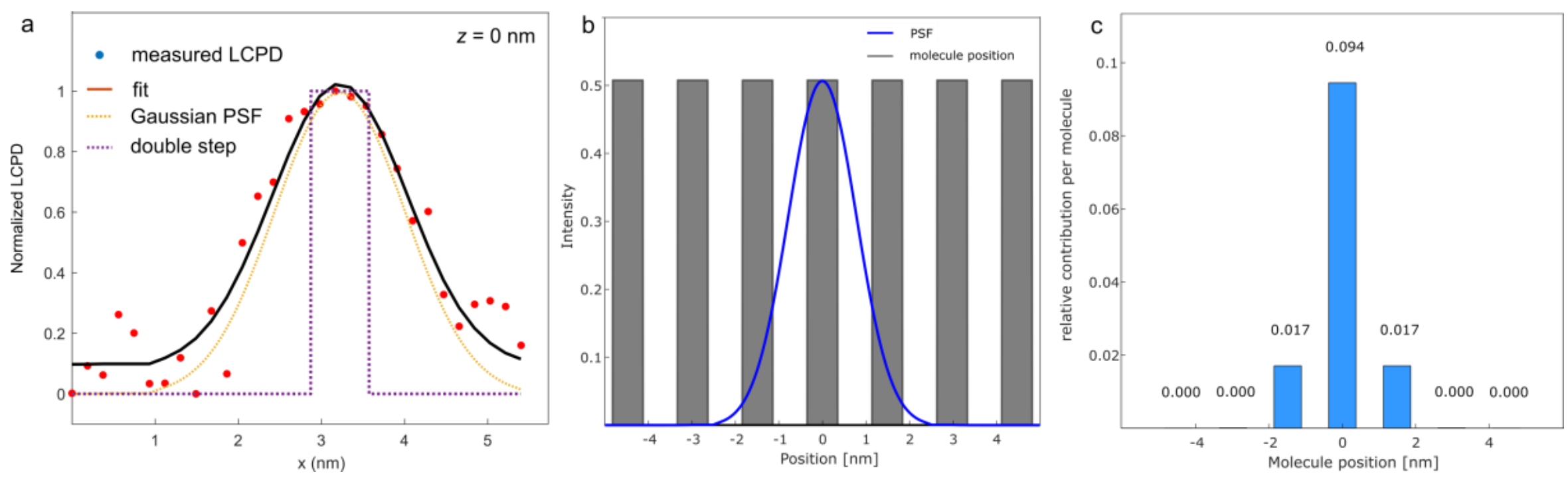

**Fig. S6.** Deconvolution of the measured LCPD in a $C_{60}$ stripe phases, shown in Fig. 4b. (a) Double step functions with a width of 0.7 nm have been convolved with a gaussian point spread function (PSF) to fit the experimental data taken along the profile shown in Fig. 4. (b) The Gaussian PSF from the fit ($FWHM_{PSF}$ = 1.85±0.15 nm) was used to determine the contribution of the molecules in a 1-dimensional stripe with center-center distances of 1.47 nm. (c) Average contribution of the individual molecules to the total signal.

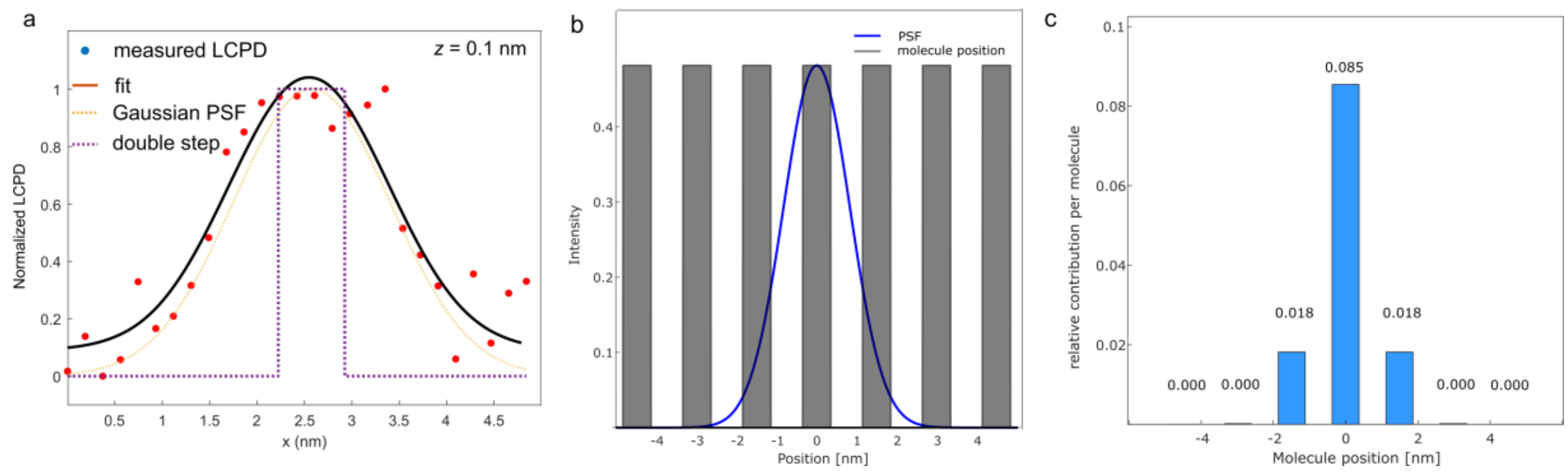

**Fig. S7.** Deconvolution of the measured LCPD in a $C_{60}$ stripe phases, along the profile in Fig. 4b after retracting the tip by 0.1 nm. (a) Double step functions with a width of 0.7 nm have been convolved with a gaussian point spread function (PSF) to fit the experimental data taken along the profile shown in Fig. 4. (b) The Gaussian PSF from the fit ($FWHM_{PSF}$ = 1.95±0.23 nm) was used to determine the contribution of the molecules in a 1-dimensional stripe with center-center distances of 1.47 nm. (c) Average contribution of the individual molecules to the total signal.

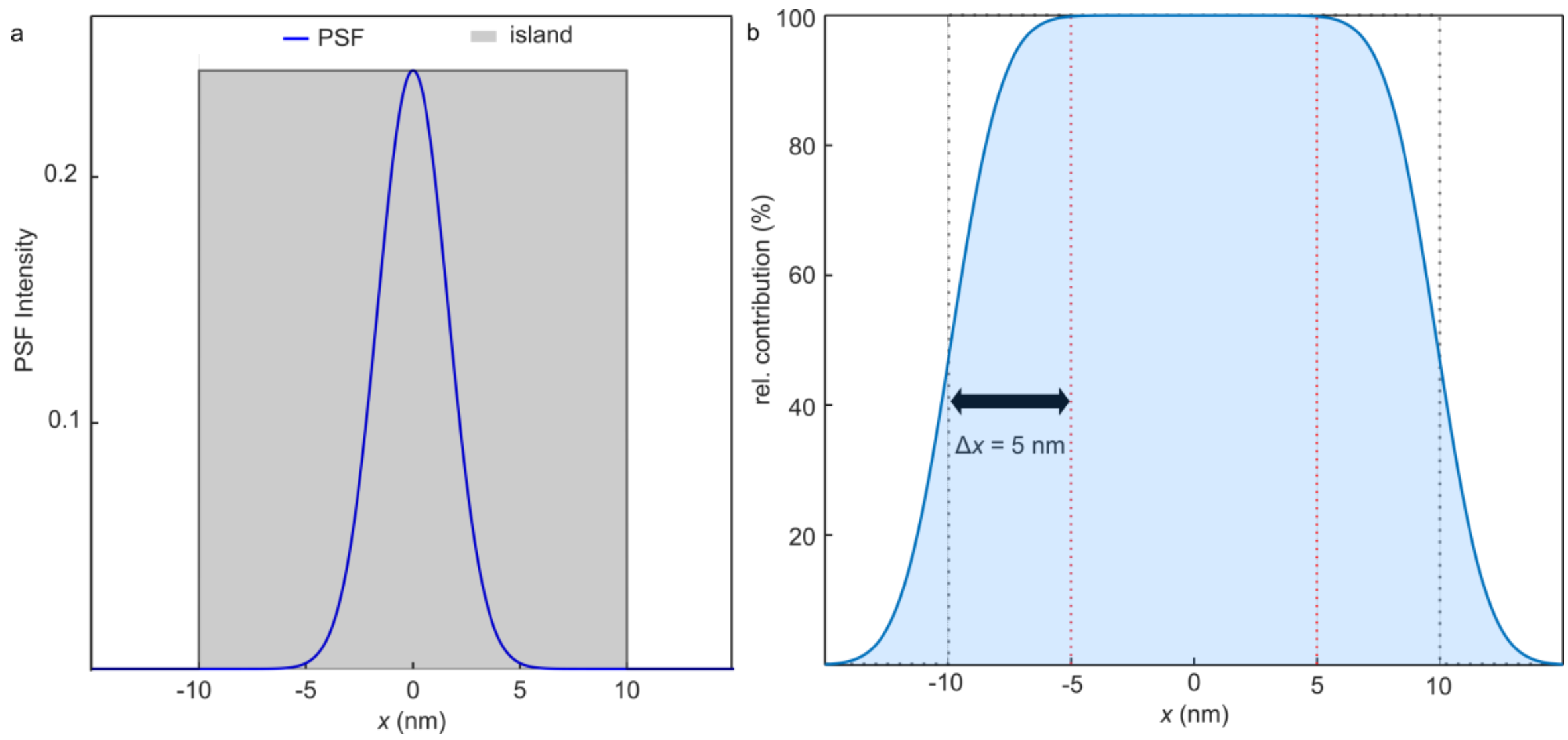

**Fig. S8.** Deconvolution of the measured LCPD for a 20 nm island. (a) The Gaussian PSF from the fit in Fig. S3a (FWHM = 3.86 nm) was used to determine the contribution of a 20 nm island. (b) Contribution of the island as a function of the tip position.

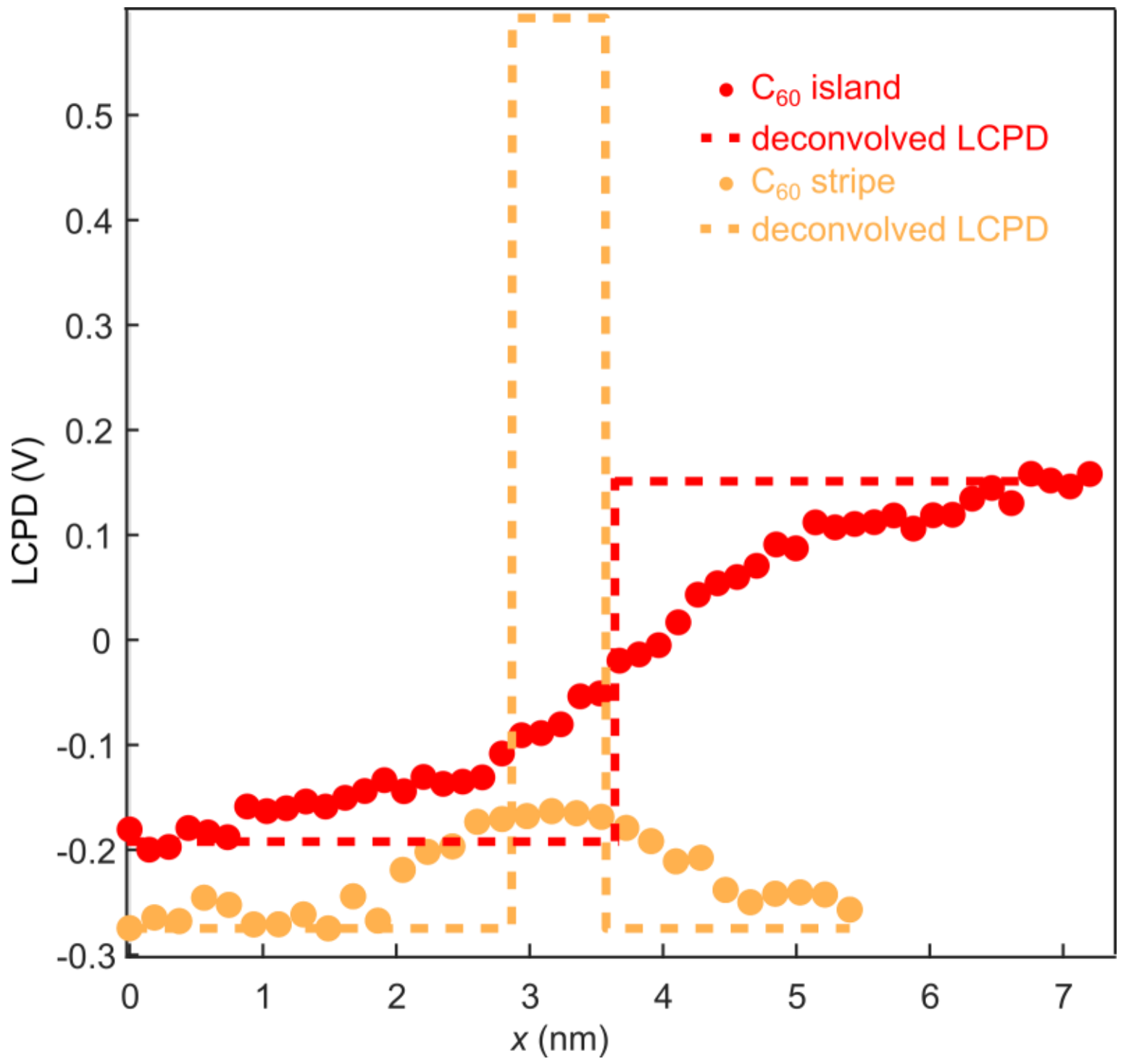

**Fig. S9.** Deconvolved LCPD for the $C_{60}$ stripe phase and island. LCPD acquired with constant height feedback along the path indicated by the arrows in Fig. 4a: across the edge of a $C_{60}$ island (red) and a $C_{60}$ stripe (orange). Feedback opened at $I$ = 10 pA, $V$ = 0.25 V, $\Delta z$ = 0.0 nm, $A$ = 0.3 nm.

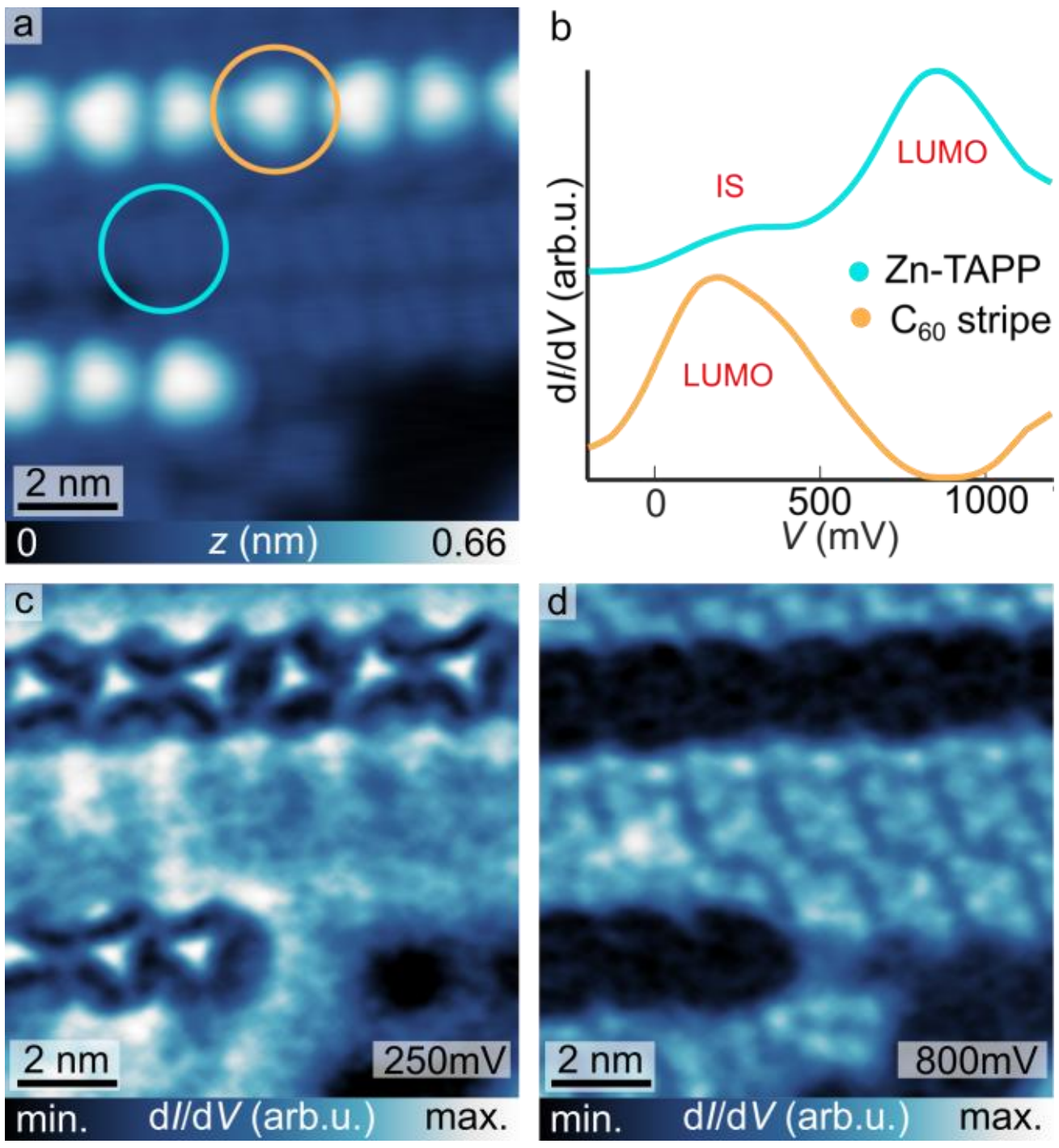

**Fig. S10.** (a) Constant-current STM image of the stripe phase. (b) d*I*/d*V* spectra acquired on $C_{60}$ and Zn-TAPP. (c-d) Constant-current d*I*/d*V* maps acquired at *V* = 250 mV and 800 mV. Measurement parameters: (a) tunneling current *I* = 50 pA, sample bias voltage *V* = 0.25 V. (b) Feedback opened at *I* = 50 pA, *V* = 1.20 V (c) *I* = 50 pA, *V* = 0.25 V. (e) *I* = 50 pA, *V* = 0.80 V. Sample bias voltage modulation $V_{mod}$ = 0.04 $V_{rms}$.

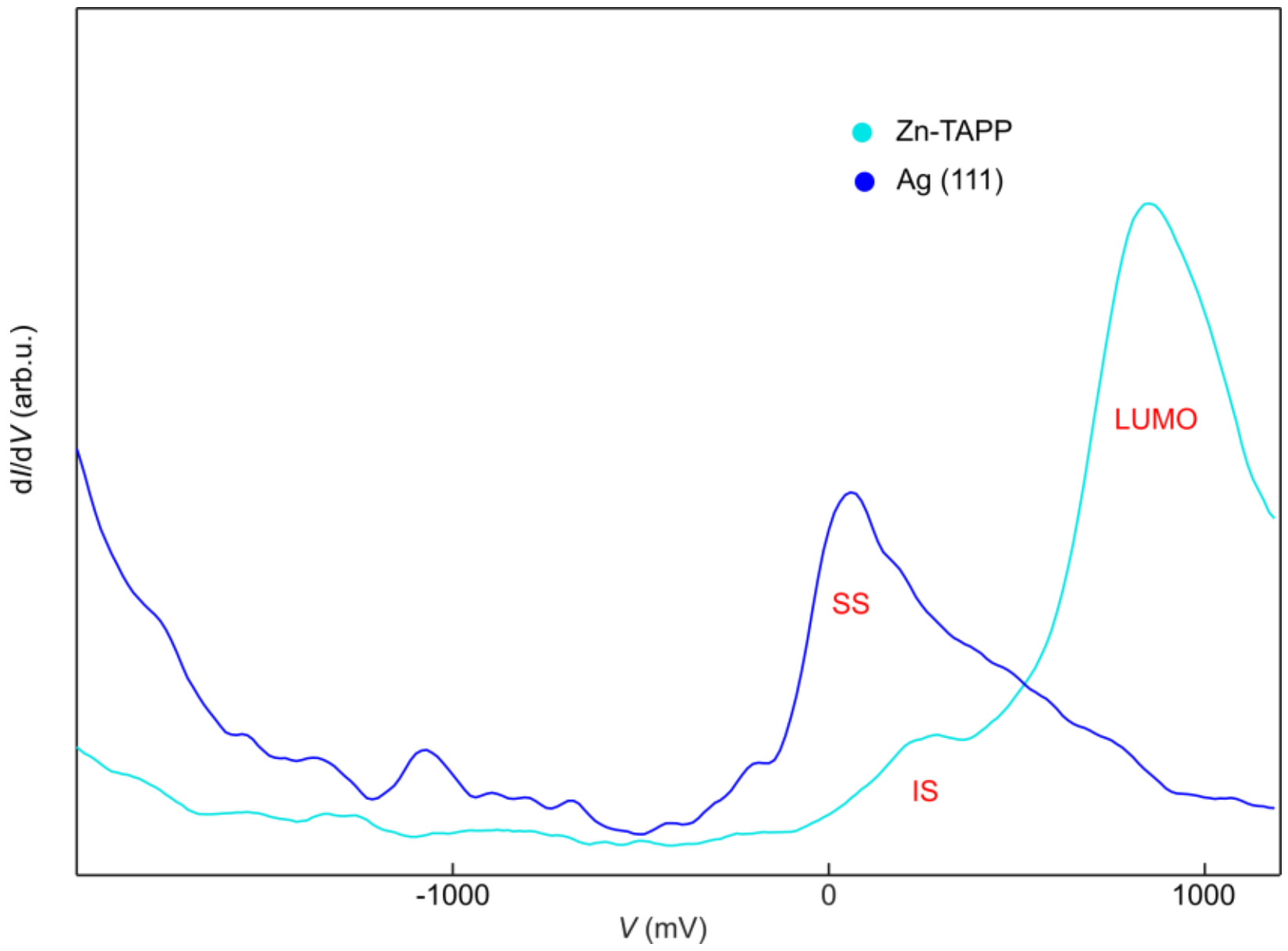

**Fig. S11.** d*I*/d*V* spectra acquired on Zn-TAPP and the Ag(111) substrate. Surface state (SS) of Ag(111), the LUMO and an interface state (IS) on Zn-TAPP are indicated. Measurement parameters: Feedback opened at *I* = 60 pA, *V* = 0.80 V. Sample bias voltage modulation $V_{mod}$ = 0.04 $V_{rms}$.

**Table S1:** Fit parameters for the spectra shown in Figure 5b,c,e,f.

| Spectrum | $C_{60}$ dim 3-fold | $C_{60}$ bright 2-fold | $C_{60}$ bright 3-fold | $C_{60}$ stripe |
|---|---|---|---|---|
| Amp. 1 | 1.259 | 2.215 | 1.950 | 4.822 |
| Center 1 | 0.135 | 0.233 | 0.331 | 0.149 |
| Width 1 | 0.102 | 0.109 | 0.144 | 0.145 |
| Amp. 2 | 3.634 | 1.910 | 1.215 | 0.000 |
| Center 2 | 0.471 | 0.431 | 0.600 | 0.500 |
| Width 2 | 0.152 | 0.159 | 0.103 | 0.100 |
| Amp. 3 | 1.405 | 1.534 | 0.939 | 3.988 |
| Center 3 | 0.810 | 0.727 | 0.788 | 0.420 |
| Width 3 | 0.119 | 0.180 | 0.153 | 0.143 |
| Amp. 4 | 2.294 | 3.264 | 2.534 | 1.733 |
| Center 4 | 1.286 | 1.281 | 1.242 | 1.239 |
| Width 4 | 0.180 | 0.180 | 0.168 | 0.163 |
| Baseline | 1.038 | 0.517 | -0.045 | 1.790 |

**Table S2:** Estimated degree of charge transfer from the spectra shown in Figure 5b,c,e,f. Degree of charge transfer estimated from the areas of three gaussian functions below 0 V (according to equation 1)

| Spectrum | $C_{60}$ dim 3-fold | $C_{60}$ bright 2-fold | $C_{60}$ bright 3-fold | $C_{60}$ stripe |
|---|---|---|---|---|
| Area 1 | 0.321 | 0.608 | 0.704 | 1.755 |
| Area 2 | 1.388 | 0.763 | 0.315 | 0.000 |
| Area 3 | 0.418 | 0.692 | 0.359 | 1.434 |
| Total Area | 2.127 | 2.063 | 1.378 | 3.189 |
| % Below Zero | 1.462 | 0.621 | 0.554 | 8.437 |
| $n_{e^-}$ / $C_{60}$ | 0.088 | 0.037 | 0.033 | 0.506 |

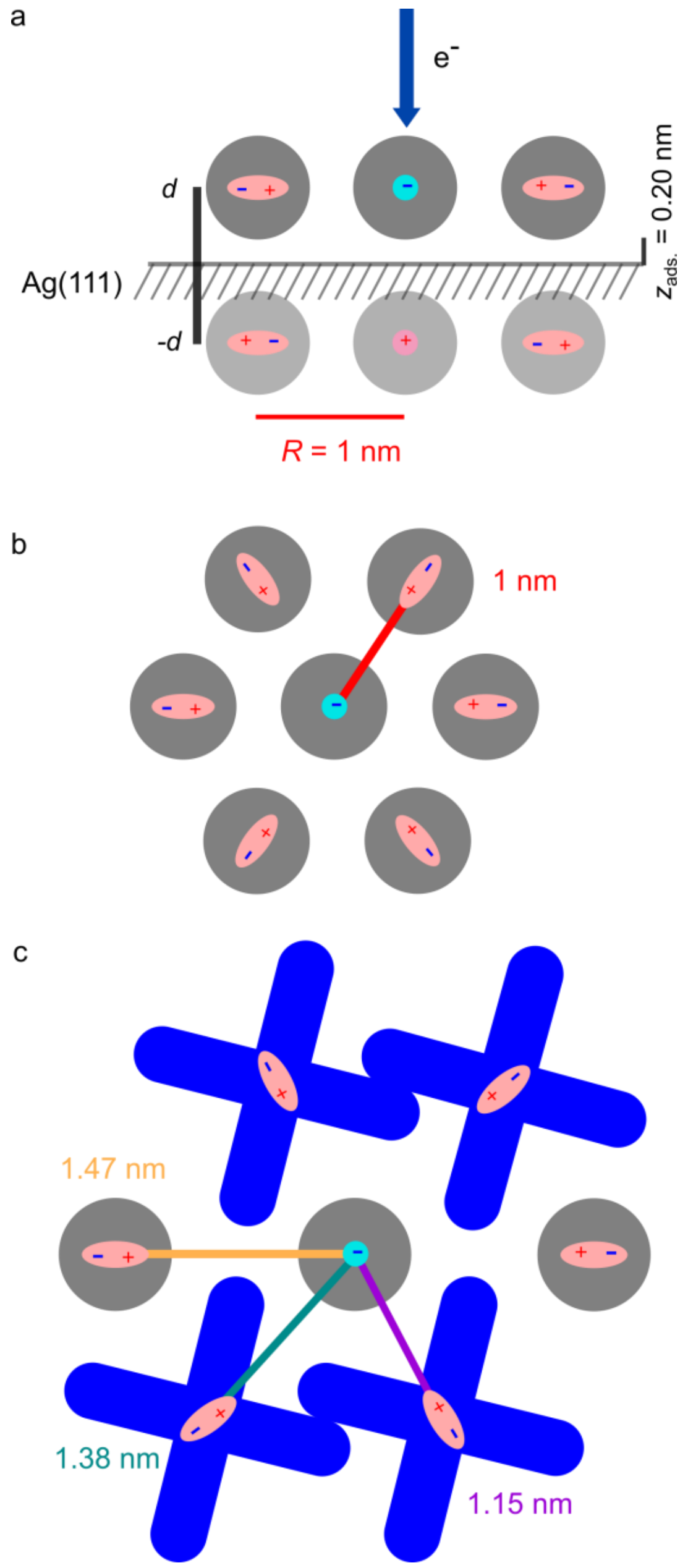

**Fig. S12.** (a) Schematic drawing of the image charge screening in the underlying Ag(111) substrate and environmental screening. Sketches of the $C_{60}$ molecular arrangements in close packed islands (b) and in the stripe phase (c).

The gas-phase polarizabilities of $C_{60}$ and Zn–TAPP were computed with DFT as implemented in ORCA 6.1.0[73-75]. Prior to the polarizability calculations, the molecular geometries were optimized using the hybrid B3LYP functional[76,77] together with the def2-TZVP basis set[78] and the def2/J auxiliary basis[79,80], employing the RIJCOSX approximation[80] and Grimme's D3 dispersion correction with Becke–Johnson damping (D3BJ[74]). The static polarizabilities were then evaluated at the B3LYP level using the diffuse-augmented def2-TZVPD basis set[78,81,82], in combination with the def2/J auxiliary basis, the RIJCOSX approximation, and the D3BJ dispersion correction.

To estimate the environmental screening – including the image charge correction for the polarizabilities[67] – in the two-dimensional island, we treated the six nearest $C_{60}$ molecules as the local environment. Regarding the one-dimensional stripe phase, we considered that a $C_{60}$ has four Zn-TAPP molecules and two $C_{60}$ as its nearest neighbors. To take the presence of the metallic substrate into account, the anisotropic polarizabilities obtained from DFT calculations (Table S3) were corrected using equation 4. For the Zn-TAPP molecules placing a point charge in the center of the molecule serves as a reasonable approximation (Fig. S12). We assumed an adsorption height ($z_{ads.}$) similar to that of Zn(II) 5,10,15,20-tetraphenylporphyrin (Zn-TPP), which was previously determined by DFT calculations (Table S3).[83] Owing to the near-spherical nature of $C_{60}$ and the polarization effects introduced by the substrate,[46] treating the system with a simple point-charge model is not reliable. According to previous works on $C_{60}$, when using point charges they should be placed at the bottom of the molecule.[84] However, due to the small adsorption heights of $C_{60}$ on Ag(111),[42] further complications arise: when the charge and image charge get too close to each other the function diverges (Fig. S13). Importantly, even when placing the point charge at the center of the $C_{60}$ molecule, the environmental screening – calculated according to equation 3 – is similar for both phases (Fig. S14). Furthermore, in this simplified model the image charge screening ($\Delta E_{pol.(sub.)}$) is the dominant contribution

**Table S3.** Gas-phase polarizabilities and adsorption heights ($z_{ads.}$).

| **/** | **Zn-TAPP** | **$C_{60}$ dim** |
|---|---|---|
| $\alpha_0$ (a.u.) | 1024 | 549 |
| $z_{ads.}$ (nm) | 0.33[83] | 0.20[41] |

(Fig. S14).

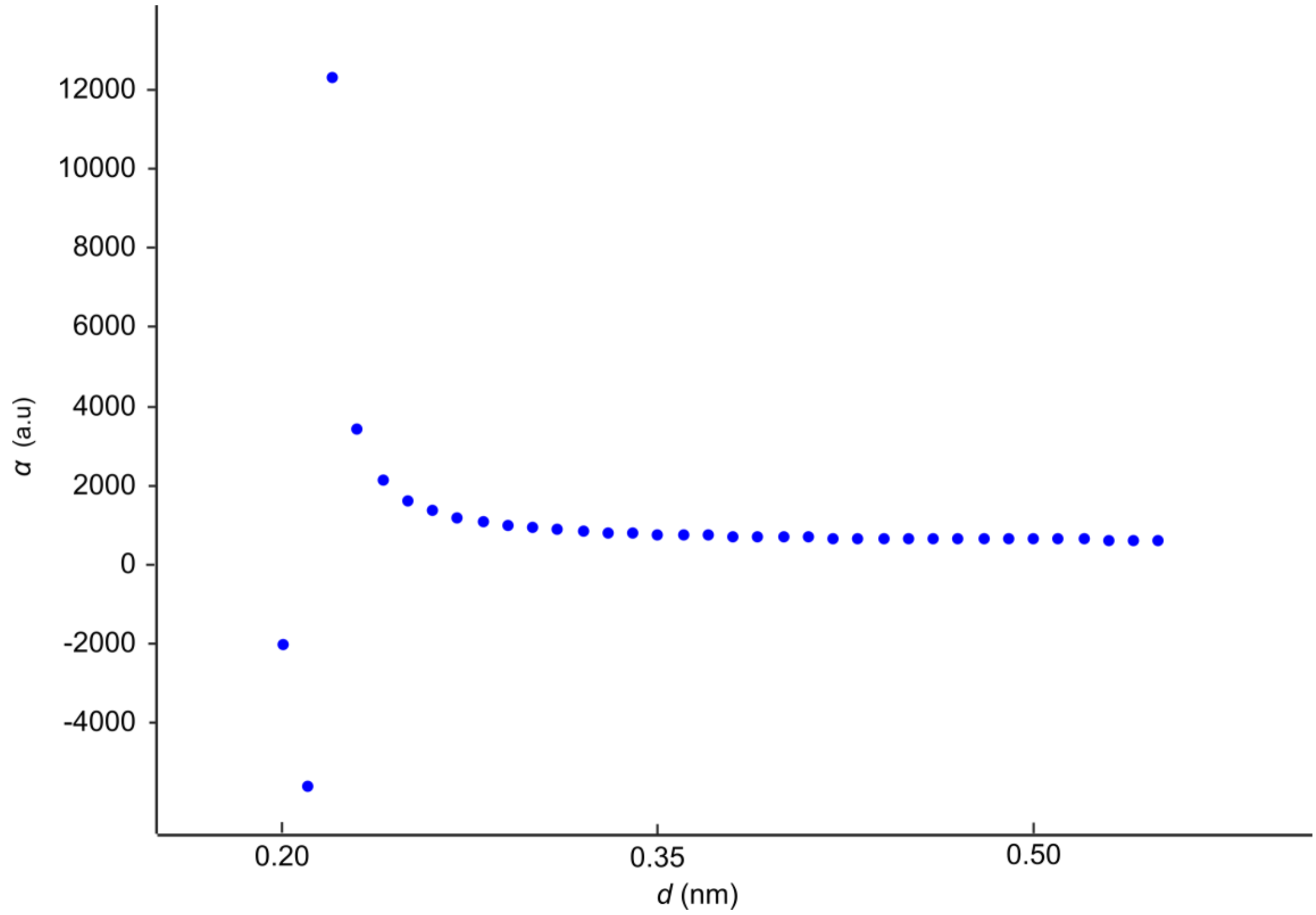

**Fig. S13.** Distance dependence of the corrected polarizability ($\alpha_{corr.}$) for a point charge.

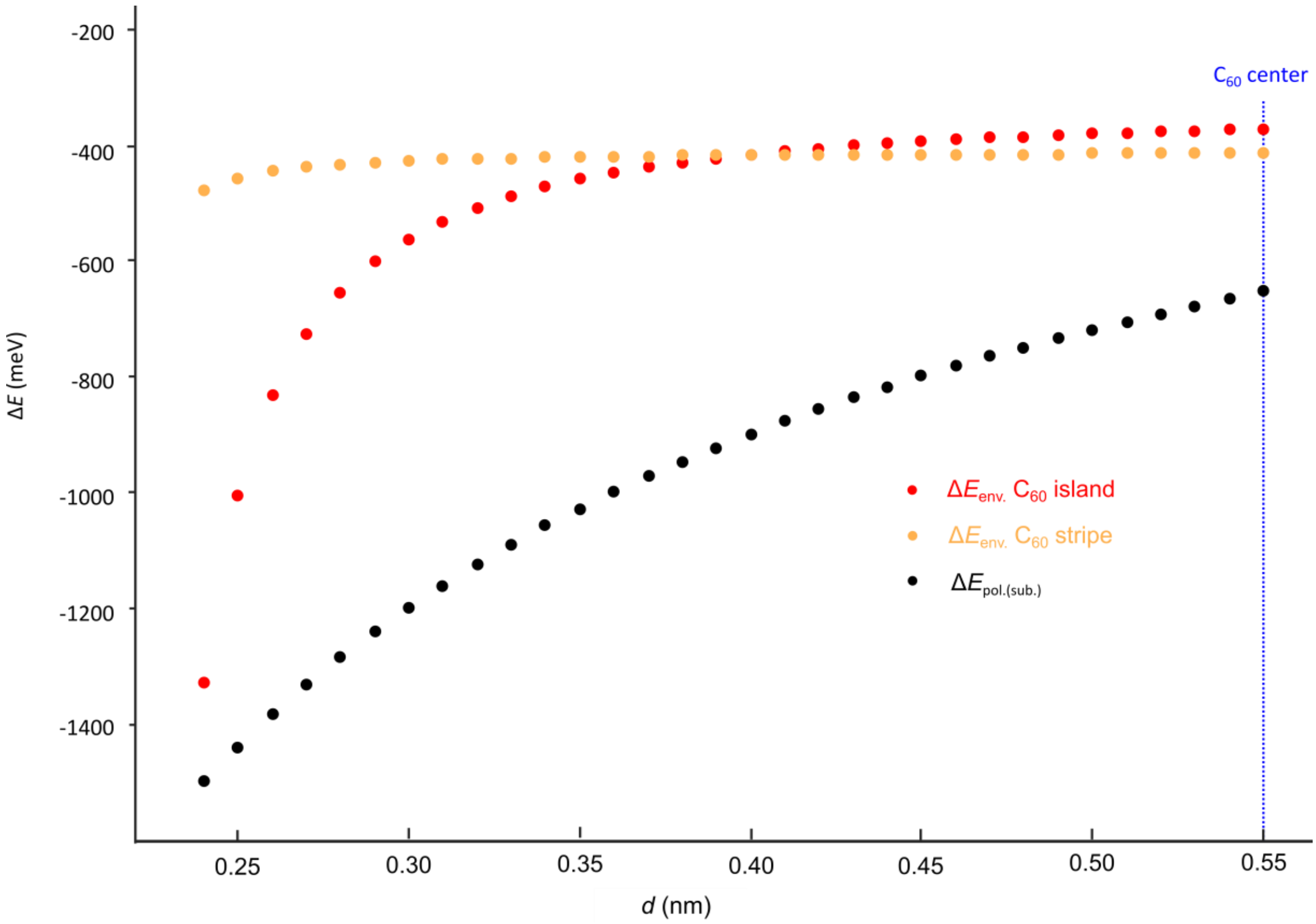

**Fig. S14.** Distance dependence of image charge screening ($\Delta E_{pol.(sub.)}$) and the environmental screening ($\Delta E_{env.}$) for a point charge the six nearest $C_{60}$.

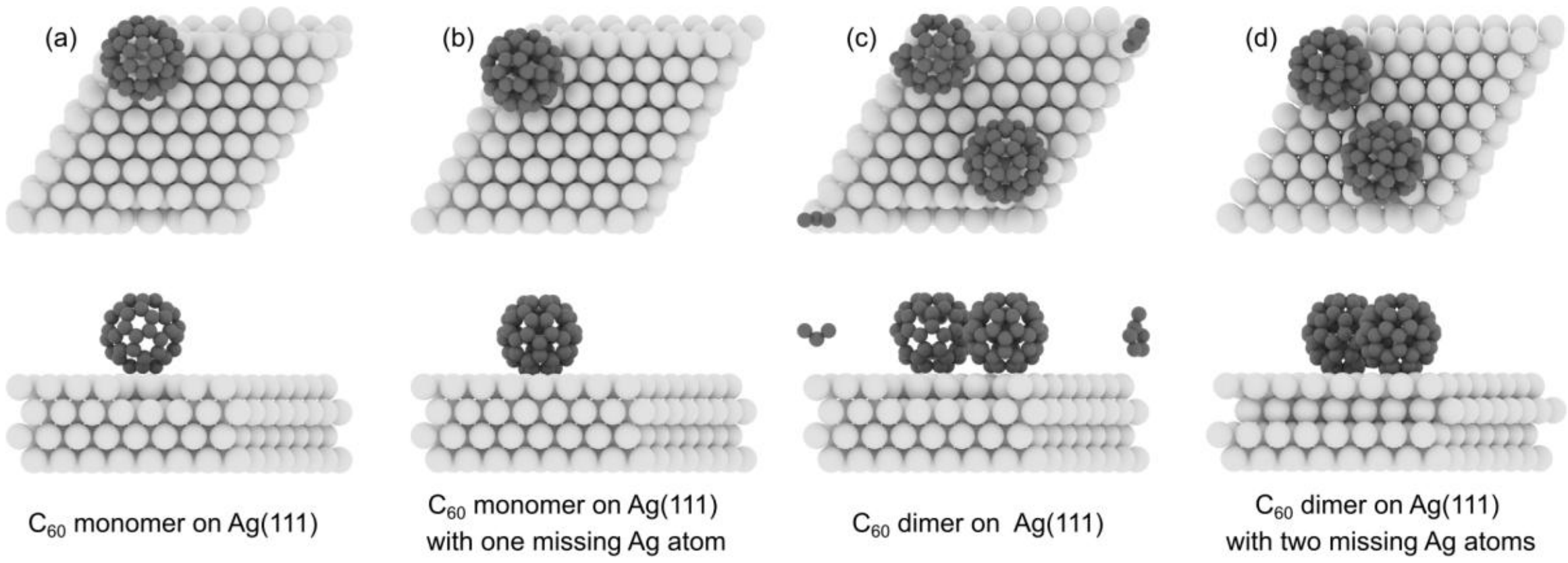

**Fig. S15.** DFT geometry optimized structures for $C_{60}$ monomers and $C_{60}$ dimers on 4 layers of Ag(111). (a,c) $C_{60}$ on unreconstructed Ag(111). (b,d) Single Ag atom vacancies underneath the $C_{60}$ molecules.

**Table S4.** Charges and adsorption energies from DFT calculations. For adsorption energies, results with two different energy cut-offs are provided, see the text.

| **Model** | **$C_{60}$ monomer unreconstructed** | **$C_{60}$ monomer reconstructed** | **$C_{60}$ dimer unreconstructed** | **$C_{60}$ dimer reconstructed** |
|---|---|---|---|---|
| $q$ | –0.65 | –0.74 | –0.59/–0.59 | –0.73/–0.73 |
| $E_{ads}$ / eV ($E_{cut}$ = 600 eV) | 2.9 | 3.0 | 5.5 | 6.2 |
| $E_{ads}$ / eV ($E_{cut}$ = 300 eV) | 2.9 | 3.1 | 5.6 | 6.3 |